\documentclass[a4paper,fleqn,usenatbib]{mnras}

\usepackage{newtxtext,newtxmath}

\usepackage[T1]{fontenc}
\usepackage{ae,aecompl}

\usepackage{graphicx}	
\usepackage{amsmath}	
\usepackage{amssymb}	
\usepackage{mathrsfs}
\usepackage{graphicx}
\usepackage{cases}
\usepackage{epsfig}
\usepackage{multirow}
\usepackage{dcolumn}
\usepackage{bm}
\usepackage{amssymb}
\usepackage{multirow}
\usepackage{enumitem}
\usepackage{enumerate}
\usepackage{subfigure}
\usepackage{color}

\graphicspath{{figure/}}

\def\prd{Phys. Rev. D}

\def\prl{Phys. Rev. Lett.}
\def\prx{Phys. Rev. X}

\def\jcap{JCAP}
\def\apj{Astrophys. J.}
\def\apjl{Astrophys. J. Lett.}

\def\mnras{Mon. Not. Roy. Astron. Soc.}
\def\apjs{Astrophys. J. Suppl. Ser.}
\def\aanda{Astron. Astrophys.}

\def\cqg{Class. Quant. Grav.}

\def\ijmpd{Int. J. Mod. Phys. D}
\def\mnrad{Mon. Not. R. Astron. Soc.}
\def \nat{Nature}

\def\be{\begin{equation}}
\def\ee{\end{equation}}

\def\bea{\begin{eqnarray}}
\def\eea{\end{eqnarray}}
\newcommand{\bes}{\begin{equation*}}
\newcommand{\ees}{\end{equation*}}
\newcommand{\beqa}{\begin{eqnarray}}
\newcommand{\eeqa}{\end{eqnarray}}

\newcommand{\lsim}{\mathrel{\hbox{\rlap{\lower.55ex\hbox{$\sim$}} \kern-.3em \raise.4ex \hbox{$<$}}}}
\newcommand{\gsim}{\mathrel{\hbox{\rlap{\lower.55ex\hbox{$\sim$}} \kern-.3em \raise.4ex \hbox{$>$}}}}
\newcommand{\Msun}{\>{M_{\odot}}}
\newcommand{\rrmag}{\>{\rm M}^{0.0}_r-5\log h}

\newcommand{\tabincell}[2]{\begin{tabular}{@{}#1@{}}#2\end{tabular}}

\title[Host galaxy groups of BBHs]{Hunting for the host galaxy groups of binary black holes and the application in constraining Hubble constant}

\author[J. Yu et al.]{
Jiming Yu,$^{1,2}$ \thanks{Email:yjm8012@mail.ustc.edu.cn}
Yu Wang,$^{1,2}$ \thanks{wywa@ustc.edu.cn}
Wen Zhao$^{1,2}$ \thanks{wzhao7@ustc.edu.cn}
and Youjun Lu$^{3,4}$ \thanks{luyj@nao.cas.cn}
\\
$^{1}$\,CAS Key Laboratory for Researches in Galaxies and Cosmology, Department of Astronomy, University of Science and Technology of China, \\Chinese Academy of Sciences, Hefei, Anhui 230026, China\\
$^{2}$\,School of Astronomy and Space Science, University of Science and Technology of China, Hefei, 230026, China \\
$^{3}$\,CAS Key Laboratory for Computational Astrophysics, National Astronomical Observatories, Chinese Academy of
Sciences, Beijing, 100101, China \\
$^{4}$\,School of Astronomy and Space Science University of Chinese Academy of Sciences, Beijing, 100049, China
}

\date{Accepted XXX. Received YYY; in original form ZZZ}

\pubyear{2020}

\begin{document}
\label{firstpage}
\pagerange{\pageref{firstpage}--\pageref{lastpage}}
\maketitle

\begin{abstract}
{ The discovery of gravitational-wave (GW) signals, produced by the coalescence of stellar-mass binary black holes (SBBHs), opens a new window to study the astrophysical origins and dynamical evolutions of compact binaries. In addition, these GW events can be treated as the standard sirens to constrain various cosmological parameters. Both issues require the host identification for these GW events, with help of the spatial resolution of GW detector networks. In this paper, we investigate the capabilities of various detector networks for identifying the SBBHs' host galaxy groups, rather than their host galaxies, which can overcome the influence of galaxies' proper motions in dark matter halos for measuring the cosmological parameters. In our analysis, the group catalog of SDSS DR7 with redshift $z\in(0.01,0.1)$ is considered as an example of the application. We find that for the second-generation (2G) detector network, the host galaxy groups of around $(0.7-6.9)$ SBBHs can be identified per year assuming all sources are $30-30\ M_{\odot}$ binaries, and that all five detectors in the network are in lock 100\% of the time. For the 3G detector network, this number becomes $(3.9-40.0)$ yr$^{-1}$. We also investigate the potential constraint on the Hubble constant $H_0$ by these GW events, if their redshift information is extracted from the candidates of host galaxy groups. We find that, by five-year's full time observations, 2G detector network is expected to give a constraint of $\Delta H_0/H_0\sim (1\%,4\%)$, which can be more than two order smaller if considering the 3G detector network.}

\end{abstract}




\section{Introduction}

\label{intro}

The discoveries of gravitational-wave (GW) signals produced by the inspiral and merger of stellar-mass binary black holes (SBBHs) and binary neutron stars (BNSs), by advanced LIGO and Virgo collaborations (LVC) \citep{abbott2016a,abbott2016b, abbott2016c, abbott2016d, abbott2017a, abbott2017b, abbott2017c, abbott2017d, abbott2019a}, confirm the existence of GWs, and provide a great support for the Einstein's general relativity. In addition, these discoveries also reveal the existence of a large cosmic population of SBBHs. In consequence, the astrophysical origins of these sources, as well as the related abundant stellar and dynamical physics, become one of the important tasks in the community, which could provide an crucial clue for the population synthesis models. Recently, various mechanisms, in particular the one related to the evolution of massive binary stars in galactic fields, have been proposed to produce SBBHs \citep[e.g.,][]{Rodriguez2015,Rodriguez2016,Mapelli2016,Wang2016,Sasaki2016,McKernan2018,Wang2018}. 
Since we lack of information on SBBHs \citep{Schneider2017}, it is very difficult to distinguish these models. In different mechanisms, the local environment, for instance the properties of host galaxies, host galaxy groups or host galaxy clusters, of the SBBH GW events are always different. Therefore, it is anticipated that the properties of the hosts of GW sources can be used to reveal the formation mechanisms for SBBHs and constrain the physics involved in the SBBH formation processes \citep{artale2019, artale2020, adhikari2020}. For this reason, the identifications of the host galaxies, groups or clusters of GW events by the GW and/or electromagnetic (EM) observations are essential for this issue.

As the prime candidate sources of ground-based GW detectors, the GW signals generated by the coalescence of compact binaries also open a new window for the research of cosmology. From the GW signal itself, one can measure the luminosity distance of the binary coalescence without having to rely on a cosmic distance ladder \citep{schutz1986}. If one can also measure independently the source's redshift, this kind of GW events can be treated as standard siren to measure various cosmological constant \citep{Sathya2010,zhao2011,yan2020,wang2020} or calibrate the standard candles \citep{zhao2017,sathya2019}. The first application of this method was carried out by LVC. On 17 August 2017, advanced LIGO and Virgo detectors observed a GW event produced by a BNS merger (GW170817) \citep{abbott2017c}. Less than two seconds after the merger, a $\gamma$-ray burst (GRB170817A) was observed \citep{abbott2017d, goldstein2017, savchenko2017}. They were emitted from the same sky area. Subsequently, the host galaxy of this event were confirmed to be NGC4993 \citep{coulter2017, soares2017, valenti2017, arcavi2017, tanvir2017}. Combining with the GW observation, LVC derived the first constraint on Hubble constant from GW standard sirens $H_{0}=70.0^{+12.0}_{-8.0}\ \mathrm{km}\ \mathrm{s}^{-1}\ \mathrm{Mpc}^{-1}$ with 68.3\% confidence level \citep{ligo2017}. As mentioned, in this method, the crucial role is to determine the redshift of the GW events. For the GW signals produced by BNSs or NSBHs, the abundant EM signals will be emitted at the same time, which are expected to be detected and used to hunt for the EM counterparts of the GW events. Thus, the redshift could be fixed from the EM observations. The GW170817 event is a typical example of this case. However, for SBBH mergers, it seems difficult to find their EM counterparts. Although many theories predict the existence of EM counterparts \citep{mosta2010, moesta2012, schnittman2011, dotti2006}, at least until now we have not observed it. Therefore, we need to use other methods to find the redshift information of the GW sources \citep{taylor2012,messenger2012,farr2019,oguri2016, mukherjee2018}. Among them, the most natural way is to identify the host galaxies, groups or clusters of GW events with help of the spatial resolution of GW detectors \citep{pozzo2012, chen2017, fishbach2019, soares2019, palmese2019, abbott2019b, gray2019}.

For a given GW event of SBBH, observed by the GW detector network, the possibility of the host galaxy (or host galaxy group, host galaxy cluster) identification is determined by the ability to localize the source in three dimensions, i.e. the detector's angular resolution $\Delta \Omega_s$ and the depth of localization quantified by the luminosity distance $d_{L}$ estimation. As well known, in addition to the signal-to-noise ratio (SNR) of the detection, the value of $\Delta \Omega_s$ also depends on the detector's spatial distribution of the network, in particular the values of the detector spacing \citep{wen2010}. On the other hand, the uncertainty $\Delta d_L$ mainly depends on the value of SNR and the inclination angle $\iota$ of the GW event, due to the strong correlation between the parameters $\iota$ and $d_L$ in the data analysis \citep{zhao2018}. In the near future, two more detectors will be built. They are KAGRA in Japan \citep{abbott2018a} and LIGO-India \citep{unni2013}. Together with two advanced LIGO detectors and advanced Virgo, they will form a network of five second-generation (2G) ground-based laser interferometer GW detectors. Looking further ahead, there are two leading proposals under consideration for the design of the third-generation (3G) GW detectors. One is the Einstein Telescope (ET) in Europe \citep{punturo2010, abernathy2011} and the other is Cosmic Explorer (CE) in the U.S. \citep{abbott2017f, dwyer2015}. These two detectors may come on line in 2030s. Beyond this, in 2015, two 8-km LIGO-type detectors were proposed in \cite{blair2015} and \cite{howell2017}. The one would be in Australia (denoted as AIGO detector in \cite{blair2015} and the other one would be in China. With these more advanced detectors, we will be able to detect more GW events, and the ability to locate GW sources will be greatly improved. In this work, we will consider various networks, consisting of the mentioned GW detectors, in the 2G and 3G eras. We should also mention that, in addition to these detectors, various upgrades have been proposed for the advanced LIGO detectors \citep{LIGOA+} leading to the proposal for an upgrade to $A+$ in 2020 followed by a further upgrade to LIGO Voyager which is envisioned to be operational around 2025. Since the noise level of this proposal is close to that of AIGO, we will not focus on this potential detector in this work.

If the spatial position of the GW event can be well determined by GW observation, in the universe with fixed cosmological parameters, the candidates of its hosts are possible to be identified at a typical distance of 1 Gpc observed by the 2G detector networks, the values of $\Delta\Omega_s$ can range from several to hundreds of square degrees. Since such a sky patch can contain thousands of galaxies, the chances of identifying the host galaxy of a GW event from galaxy catalog can be dismal. The typical examples of the application were carried out in \cite{fishbach2019} and \cite{soares2019}, where the authors used the localizations of GW events GW170817 and GW170814 respectively to identify the candidates of their host galaxies. For each event, one can only derive the a number of candidates of its host galaxies, which follows a wide distribution of the event's redshift value. Combining with the measurement of the luminosity distance, the authors derived the loose constraint on the Hubble constant, which are $H_{0}=76^{+48}_{-23}\ \mathrm{km}\ \mathrm{s}^{-1}\ \mathrm{Mpc}^{-1}$ and $75^{+40}_{-32}\ \mathrm{km}\ \mathrm{s}^{-1}\ \mathrm{Mpc}^{-1}$ for GW170817 and GW170814, respectively.

In order to overcome the difficulty of host galaxy identification as mentioned above, a nature choice is to consider the larger structures in the universe, i.e. identifying the host galaxy groups or host galaxy clusters of GW events \citep{macleod2008}. According to the current scenario of structure formation, galaxy groups are associated with cold dark matter halos. Therefore, the halo-based galaxy group stands for the larger physical structure, than the galaxy, in the universe. In this paper, we shall extend the idea to identify the host groups of SBBHs by utilizing the GW observations. In comparison with the host galaxies, the host groups of the GW events are easier to be identified due to their larger spatial sizes, and the properties of host groups can be used to study the formation circumstances of the SBBHs. In this paper, we shall focus on two problems: (1) For the given cosmological parameters, how well can the host groups of SBBHs be identified by the GW observations in the eras of 2G and 3G GW detectors? (2) How well can the Hubble constant $H_0$ be measured, by adopting the redshift information of SBBHs provided by the identified host groups? In the calculation, we use the galaxy group catalog of SDSS DR7 \citep{abazajian2009} as the example. The groups we used are obtained in a broad dynamic range and include systems of isolated galaxies \citep{yang2007}. We assume that SBBH mergers in each group, and then use the Fisher information matrix \citep{wen2010} to estimate the three-dimensional spatial position of the source. After that, we count the number of groups $N_{in}$ in the location, which represents the strength of the detector's ability to directly locate the host groups. Then, for each detector network, we use the assumed SBBH mergers to measure the Hubble constant by the Bayesian method. For the determination of cosmological parameters, comparing with the way related to the host galaxies identification, the method by identifying the host groups has another two advantages: First, it can partly overcome the influence of peculiar velocities of GW events. As well known, the contribution of the peculiar velocity of the GW events leads to a bias in the inferred values of cosmological parameters, if not accounted for \citep{ligo2017}. The peculiar velocity of a host galaxy can arise from two components namely the motion of the halo due to the spatial gradients in the gravitational potential, and the virial velocity component of the host galaxy inside the halo. The former component can be well modelled in the analysis \citep{mukherjee2019}, and the latter component can be significantly reduced, if we consider the event's host group, instead of its host galaxy. Second, it can partly overcome the problems induced by the incompleteness of the galaxy catalog. Actually, some SBBHs may be at the small galaxies, which are not included in the galaxy catalog due to the limits of the telescope. Thus, their host galaxies might be mismatched in the analysis, which will bias the inferred values of cosmological parameters. However, these small galaxies are believed to be the {satellite galaxies} in the galaxy groups, which can be easier be included in the group catalog through their central galaxies. Therefore, for a given catalog, the completeness of the groups are expected to be much better than that of galaxies.

Our paper is organized as follows. In Sec. \ref{catalog} we introduce the galaxy group catalog used in this work. We introduce the detector response and the Fisher information matrix in Sec. \ref{method}. In Sec. \ref{nin}, for various detector networks, we investigate the their abilities of identifying the host groups of GW events. In Sec. \ref{hubble}, we use the Bayesian method to measure the Hubble constant and in Sec. \ref{conclusion} we summarize the main results of this work.

Throughout the paper, we adopt the standard $\Lambda$CDM model with following parameters as the fiducial cosmological model: $\Omega_{m}=0.272$,
$\Omega_{\Lambda}=0.728$, $\Omega_{b}=0.045$, $\sigma_{8}=0.807$,
$h=0.704$ \citep{komatsu2011}.

\section{Observational Data: The SDSS Group Catalog}
\label{catalog}
In this paper, as an example of the application of our method, we consider the SDSS galaxy group
catalogs of \cite{yang2007}, constructed using the
adaptive halo-based group finder of \cite{yang2005}, here
updated to Data Release 7 (DR7). In the construction all galaxies
from SDSS DR7 in the Main Galaxy Sample with redshifts in the range
$0.01 \leq z \leq 0.20$ and with a redshift completeness  $f_{\rm edge} >  0.7$
are selected. The  magnitudes and colors of  all galaxies  are based
on the  standard SDSS Petrosian technique \citep{petrosian1976, strauss2002}, and have been corrected for galactic  extinction \citep{schlegel1998}. The absolute magnitude of all galaxies
in band w (r and g band) are computed using
\begin{equation}
M_w^{0.1}-5\log h = m_w-DM(z)-K_w^{0.1}-E_w^{0.1},
\end{equation}
where $DM(z)=5 \log [D_{L}/(h^{-1}\mathrm{Mpc})]+25$ is the bolometric
distance modulus calculated from the luminosity distance $D_L$,
$K$-corrections $K_w^{0.1}$ are corrected to $z=0.1$, using the
method described in \cite{blanton2003a, blanton2003b}, and the evolution
corrections $E_w^{0.1}$ are corrected to $z=0.1$ by
$E_w^{0.1}=A_w(z-0.1)$ with $A_w=4.22, 2.04, 1.62, 1.61, 0.76$ for
corresponding band u,g,r,i,z \citep{blanton2003a}. Then, using the
relations between stellar mass-to-light ratio and $(g-r)$ color from
\citep{Bell et al.(2003)}, stellar masses for all galaxies are computed
using
\begin{eqnarray}
\log\left[{M_* \over h^{-2}\Msun}\right] & = & -0.306 + 1.097
\left[(g-r)^{0.0}\right] - 0.10 \nonumber\\
& & - 0.4(\rrmag-4.64)\,,
\end{eqnarray}
where $(g-r)^{0.0}$ and $\rrmag$  are the  $(g-r)$ color  and
$r$-band magnitude $K+E$ corrected to $z=0.0$, $4.64$ is the
$r$-band magnitude of the Sun in the AB  system \citep{blanton2007}, {\color{black}and the $-0.10$ term  effectively implies  that we adopt a \cite{kroupa2001} IMF \citep{borch2006}.}

It is worth mentioning that this group finder is optimized to assign galaxies into groups according to their common dark matter halos. Furthermore, for each group in the catalog, a dark matter halo mass $M_h$ can be
obtained  based on  the ranking of  the characteristic group stellar
mass, which is total  stellar mass of  all group members with
$r$-band absolute magnitudes brighter than $-19.5$. As indicated
by \cite{yang2007}, the estimate for groups' host halo mass is
equally applicable to groups spanning the entire range in richness,
which is beneficial for detecting the host groups of the
GW sources. Detailed tests using mock galaxy
redshift surveys have demonstrated that the group masses thus
estimated can recover the true halo masses with a 1-$\sigma$
deviation of  $\sim 0.3$ dex \citep{yang2005, yang2007}. Groups that
suffer severely from edge effects (about 1.6\% of the total) have
been removed from the catalog. The center of each cluster is
defined to be the stellar mass - weighted average of the positions
and redshifts of member galaxies brighter than $-19.5$.

In our analysis, we use massive groups with halo mass
$\log_{10}({M}_\mathrm{halo}/(\Msun\ h^{-1}))>12$. Similar with the case of identifying the host galaxies \citep{fishbach2019}, we neglect the possibility of the lighter groups becoming hosts, in order to ensure the catalogue's completeness.
Based on the halo mass, we can also estimate
the effect of the spatial distribution of the galaxies in the groups,
which is limited by the virial radius of the the group's host
halos. The virial radius of the galaxy groups' host halos can be computed using,
\begin{equation}
R_h={\left(\frac{3M_\mathrm{halo}}{4\pi\delta\bar{\rho}}\right)}^{1/3},
\end{equation}
where $\bar{\rho}$ is the average cosmic density {\color{black}and $\delta=180$ means that dark matter halos are defined as having an overdensity of 180}. {\color{black}Tables \ref{tab1} and \ref{tab2} show that the member galaxies of a dark matter halo are distributed in a small space, even for halos within the largest mass range $>10^{14.5}M_\odot$ or with the galaxy members more than 5, the effect of the halo virial radius on the redshift uncertainty of the member galaxies is only about {\color{black}$10^{-4}$.}}

\begin{figure}
\centering
	\includegraphics[width=8cm]{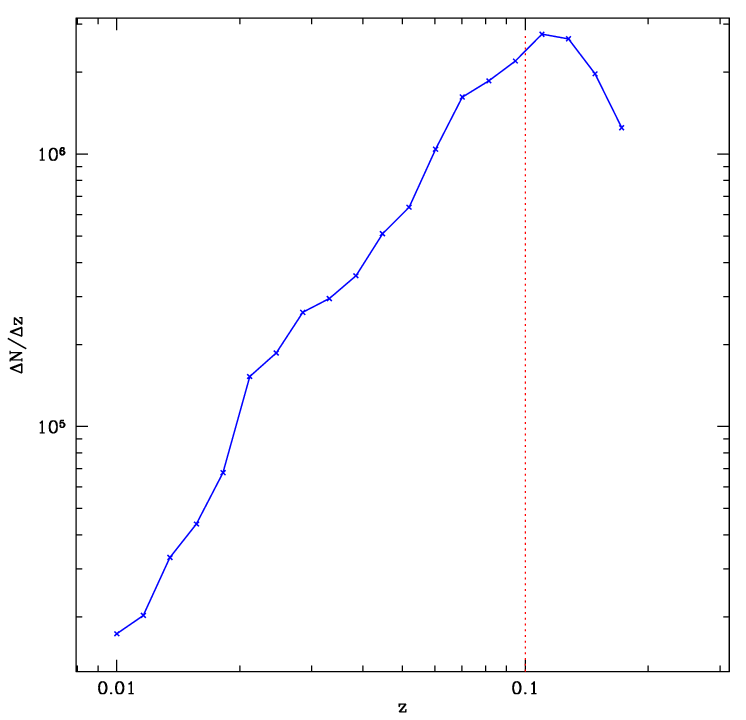}
\caption{The number density distributions of galaxy groups with respect to the redshift $z$. Here, the galaxy groups are separated into 20 redshift bins, which are set by $\log \Delta z=(\log z_{max}-\log z_{min})/20$, $z_{min}=0.01$, $z_{max}=0.2$, and $\Delta N$ is the galaxy group number in the redshift bin $\Delta z$.}
\label{figuregroup2}
\end{figure}

The incompleteness effect, due to the survey geometry, needs to be accounted
for. A group whose projected area straddles one or more survey edges may have members that fall
outside of the survey, thus causing an incompleteness, which affects
the accuracy of the detection for the host groups of the
GW sources. To avoid this incompleteness, we select the groups with the survey
edge $f_{\rm edge} > 0.7$ , which represents the volume of the group
that lies within the survey edge \citep{yang2007}. Another incompleteness effect is caused by the observational capabilities of the telescope. In the higher redshift region, limited by maximum apparent magnitudes of the targets in SDSS data, some galaxies and galaxy groups might be absent in the catalog. In Fig. \ref{figuregroup2}, we plot the number density distribution of galaxy groups with respect to the redshift $z$. Based on the cosmological principle where the galaxy group distribution at the large-scale is homogeneous and isotropic, the number density in the complete group catalog should satisfy the relation $dN/dz\propto z^{2}$ with the linear approximation, {\color{black}{that means $\log(\Delta N/\Delta z)\propto \log z $}.
}From this figure, we find that this relation is satisfied only in the region $z<0.12$. As a conservative consideration, in the following discussions, we only use the galaxy groups in the low-redshift region with $z<0.1$. The density of groups is about $2\times10^{-3}$ per Mpc$^{3}$ in this region. As a comparison, the density of galaxies in our catalogue is about $6\times10^{-3}$ per Mpc$^{3}$.


\begin{table*}
\begin{center}
\begin{tabular}{|c|c|c|c|}
\hline
$\log_{10}(M_\mathrm{halo}/(\Msun\ h^{-1}))$&Number&$\log_{10}(\overline{M}_\mathrm{halo}/(\Msun\ h^{-1}))$&$\overline{R}_{180} ( {\rm Mpc}\ h^{-1} )$\\
\hline
$>$ 14.5&530&14.71&2.048\\
\hline
14.0-14.5&3479&14.20&1.396\\
\hline
13.5-14.0&14591&13.72&0.962\\
\hline
13.0-13.5&43283&13.32&0.662\\
\hline
12.5-13.0&93268&12.75&0.456\\
\hline
12.0-12.5&132553&12.26&0.314\\
\hline
\end{tabular}
\end{center}
\caption{The distribution of the groups' halo mass in our catalog. Note that, $\overline{M}_{\mathrm{halo}}$ and $\overline{R}_{180}$ are the average mass and average virial radius of the halos in the corresponding interval respectively. }
\label{tab1}
\end{table*}

\begin{table*}
\begin{center}
\begin{tabular}{|c|c|c|c|}
\hline
Galaxy number&Number of groups&$\log_{10}(\overline{M}_\mathrm{halo}/(\Msun\ h^{-1}))$&$\overline{R}_{180} ( {\rm Mpc}\ h^{-1} )$\\
\hline
1&230399&12.76&0.414\\
\hline
2&35386&13.23&0.562\\
\hline
3&9571&13.48&0.675\\
\hline
4&4141&13.63&0.766\\
\hline
5&2258&13.74&0.834\\
\hline
$>$ 6&5949&13.98&1.028\\
\hline
\end{tabular}
\end{center}
\caption{This table shows the distribution of the groups' galaxy number in our groups. Note that, $\overline{M}_{\mathrm{halo}}$ and $\overline{R}_{180}$ are the average mass and average virial radius of the halos in the corresponding interval respectively.}
\label{tab2}
\end{table*}

\section{The detector's response and the Fisher information matrix}
\label{method}
\subsection{The detector's response}
\label{methodA}
\begin{figure}
\centering
\includegraphics[width=8cm,height=7cm]{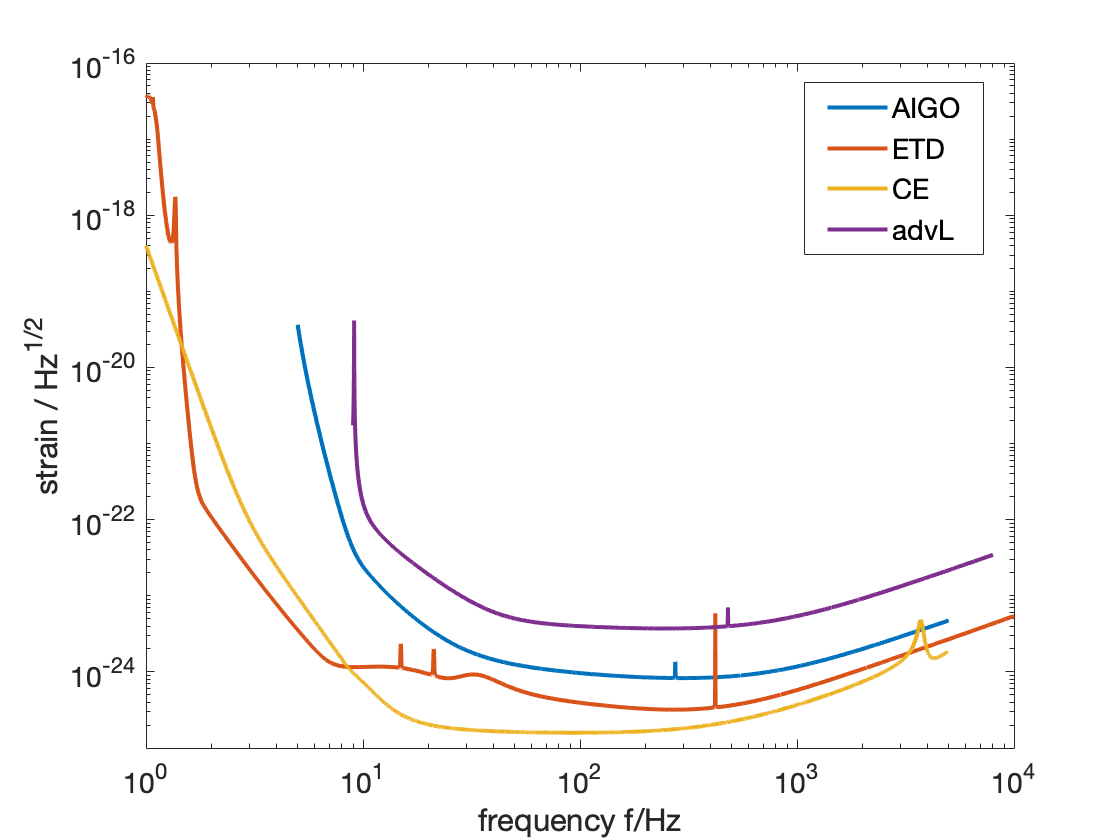}
\caption{The sensitivity noise curves of GW detectors. The purple line represents the noise level of advanced LIGO, the blue line represents that of the proposed 8-km detectors in China and Australia \citep{blair2015}, the red line represents that of ET detector \citep{punturo2010}, and the yellow line represents that of CE detector \citep{abbott2017f}.}\label{figurenoise}
\end{figure}

Let us consider a GW detector network consisting of $N_d$ detectors. We assume their spatial locations are given by the vector $\mathbf{r}_{I}$ with $I=1,2,\cdots,N_{d}$. Each detector has spatial size much smaller than the GW wavelength. For detector $\mathbf{r}_{I}$, the response to an incoming GW signal could be written as a linear combination of two wave polarizations in the transverse traceless gauge,
\be
	d_{I}(t_{0}+\tau_{I}+t)=F^{+}_{I}h_{+}(t)+F^{\times}_{I}h_{\times}(t),\ \ 0<t<T,
\label{2.1}
\ee
where $h_{+}$ and $h_{\times}$ are two polarization modes of GW, $t_0$ is the arrival time of the wave at the coordinate origin. The term $\tau_{I}$ is the time required for the waves to travel from the origin to reach the $I$-th detector at time $t$, i.e.,
\be
	\tau_{I}(t)=\mathbf{n}\cdot\mathbf{r}_{I}(t),
\label{2.2}
\ee
where $\mathbf{n}$ is the propagation direction of a wave, $t\in[0, T]$ is the time label of the wave, and $T$ is the time duration. The quantities $F^{+}_{I}$ and $F^{\times}_{I}$ are the detector's antenna beam-pattern functions, which depend on the source location $(\alpha, \delta)$, the polarization angle $\psi$, the detector's latitude $\lambda$, longitude $\varphi$, the angle $\gamma$, which determines the orientation of the detector's arms with respect to local geographical directions, and the angle between the interferometer arms $\zeta$ \citep{jaranowski1998}. Table \ref{table1} lists the parameters of interferometers used in this paper \citep{blair2015,vitale2017}. In Fig. \ref{figurenoise}, we show the sensitivity noise curves of these detectors. For LIGO (Livingston), LIGO (Handford), Virgo, KAGRA, LIGO-India, we use the noise curve of the designed level for advanced LIGO. For the proposed 8 km detectors in China and Australia, we use the noise curve given in \cite{blair2015} and \cite{howell2017}. For CE in the U.S. and the assumed CE-type detector in Australia, we use the proposed noise curve in \cite{abbott2017f} and \cite{dwyer2015}. And for ET, we consider the proposed ET-D project \citep{punturo2010, abernathy2011}.

%


The Fourier transform of the detector's response is given by
\be
	d_{I}(f)=\int^{T}_{0}d_{I}(t)e^{2\pi ift}dt.
\label{2.3}
\ee
Denoting the corresponding one-side noise spectral density of the detector by $S_{I}(f)$, we define a whitened data set in the frequency domain \citep{wen2010, zhao2018},
\be
	\hat{d}_I(f)\equiv S_{I}^{-1/2}(f)d_{I}(f).
\label{2.4}
\ee
For a detector network, Eq. (\ref{2.4}) could be written as \citep{wen2010, zhao2018}
\be
	{\bf \hat{d}}(f)=\hat{\mathbf{A}}\mathbf{h}(f)e^{-i\Phi},
\label{2.5}
\ee
where $\hat{\mathbf{A}}\mathbf{h}(f)$ is
\be
	\hat{\mathbf{A}}\mathbf{h}(f)=\left [
	\frac{F^{+}_{1}h_{+}(f)+F^{\times}_{1}h_{\times}(f)}{\sqrt{S_{1}(f)}},
	\cdots,
	\frac{F^{+}_{N_{d}}h_{+}(f)+F^{\times}_{N_{d}}h_{\times}(f)}{\sqrt{S_{N_{d}}(f)}}
	\right ]^{T},
\label{2.6}
\ee
and the detail of the waveform term $F^{+}_{I}h_{+}(f)+F^{\times}_{I}h_{\times}(f)$ could be found in \cite{ajith2008}.
$\Phi$ is a diagonal matrix of $N_{d}$ dimensions with
\be
	\Phi_{IJ}=2\pi f\delta_{IJ}(\mathbf{n}\cdot\mathbf{r}_{I}(f)+\tau_{e}(f)).
\label{2.7}
\ee
The term $\tau_{e}(f)$ represents the time delay caused by the revolution of the Earth. If we consider the effects due to the Earth's rotation and revolution during the GW burst, the quantities $F^{+}_{I}$, $F^{\times}_{I}$, $\Phi_{IJ}$ are the functions of time in the time domain, or the functions of frequency in the frequency domain, which can significantly improve the angular resolutions of the GW detectors \citep{zhao2018}. Since these effects are important only for the GW signals produced in the inspiralling stage of the compact binaries, in this paper, we apply stationary phase approximation (SPA) to derive the frequency-dependence of these functions \citep{maggiore2008, zhang2017, zhao2018}, which are given by
\begin{eqnarray}
	F^{+}_{I}(f)&=&F^{+}_{I}(t=t_{f}),\quad F^{\times}_{I}(f)=F^{\times}_{I}(t=t_{f}), \nonumber \\
	\Phi_{IJ}(f)&=&\Phi_{IJ}(t=t_{f}),
\label{2.8}
\end{eqnarray}
where $t_{f}=t_{c}-(5/256)\mathcal{M}_{c}^{-5/3}(\pi f)^{-8/3}$ \citep{maggiore2008}, with $t_{c}$ the binary coalescence time and $\mathcal{M}_{c}$ the observed chirp mass of binary system. Note that, due to the mass-redshift degeneracy in GW observation, the observed chirp mass relates to the intrinsic chirp mass $M_c$ by $\mathcal{M}_{c}=(1+z)M_c$. 
\par


\begin{table*}
\begin{center}
\begin{tabular}{|c|c|c|c|c|c|}
\hline
&$\lambda(degree)$&$\varphi(degree)$&$\gamma(degree)$&$\zeta(degree)$&abbreviation\\
\hline
LIGO (Livingston)&30.56&-90.77&243.0&90.0&L\\
\hline
LIGO (Handford)&46.45&-119.41&171.8&90.0&H\\
\hline
Virgo&43.63&10.5&116.5&90.0&V\\
\hline
KAGRA&36.25&137.18&---&90.0&K\\
\hline
LIGO-India&19.09&74.05&---&90.0&I\\
\hline
Assumed 8 km detector in Australia&-31.95&115.87&---&90.0&A\\
\hline
Assumed 8 km detector in China&29.2&120.5&---&90.0&C\\
\hline
Einstein telescope &43.54&10.42&19.48&60.0&ETD\\
\hline
Cosmic Explorer &30.54&-90.53&162.15&90.0&CE\\
\hline
Assumed CE-type detector in Australia&-31.51&115.74&0&90.0&CE\\
\hline
\end{tabular}
\end{center}
\caption{The parameters and the abbreviation of various interferometers used in our study \citep{blair2015, vitale2017}. Note that, $\gamma$ is measured counter-clockwise from East to the bisector of the interferometer arms. }
\label{table1}
\end{table*}

\subsection{Fisher information matrix}

By maximizing the correlation between a template
waveform that depends on a set of parameters and a
measured signal, the matched filtering provides a natural
way to estimate the parameters of the signal and their
errors. In this paper, we employ the
Fisher matrix approach for the parameter estimation in the GW analysis. In comparison with the MCMC analysis,
it represents a simple, method-independent, and reasonable
estimate for the detection and localization capability
for future experiments. The reliability of Fisher matrix analysis has been tested in our previous work \cite{zhao2018}.
In the case of a network with $N_{d}$ detectors, the Fisher matrix is given by \citep{wen2010}
\be
	\Gamma_{ij}=\left \langle \partial_{\theta_{i}}\hat{\mathbf{d}}\mid\partial_{\theta_{j}}\hat{\mathbf{d}}\right \rangle,
\label{2.17}
\ee
where $\theta_{i}$ are the free parameters to be estimated. For binary system we study, there are nine free parameters, $\bm{\theta}=\{\alpha,\delta,\psi,\iota,M,\eta,t_{c},\phi_{c},\log(d_{L})\}$. The angular brackets mean the inner product between two vectors,
\be
	\left \langle \mathbf{a} \mid\mathbf{b}\right \rangle = 2\sum_{i=1}^{N}\left \{\int_{-\infty}^{+\infty}[a_{i}b_{i}^{*}+a_{i}^{*}b_{i}]df\right \}.
\label{2.18}
\ee
The Cramer-Rao bound \citep{cramer1989} states that the inverse of the Fisher matrix is a lower bound on the covariance of any unbiased estimator of $\bm{\theta}$. In the case of high SNR, the covariance matrix of $\bm{\theta}$ is approximately given by the inverse of the Fisher matrix \citep{finn1992, finn1993, cutler1994},
\be
	V_{ij}=(\bm{\Gamma}^{-1})_{ij},
\label{2.19}
\ee
which is a covariance matrix with nine parameters in our analysis and one could choose the part related to the parameters $\{\alpha,\delta,\log(d_{L})\}$. Therefore, we could obtain the covariance matrix of $\{\alpha,\delta,\log(d_{L})\}$ from Fisher information matrix. The error in solid angle in the projected 2-dimensional sphere is \citep{wen2010}
\be
	\Delta{\Omega_{s}}=2\pi|\cos\delta|\sqrt{\left \langle \Delta\alpha^{2}\right \rangle\left \langle \Delta\delta^{2}\right \rangle-
	\left \langle \Delta\alpha\Delta\delta\right \rangle^{2}}.
\label{2.20}
\ee
And the total SNR for the GW signal is given by
\be
	{\rm SNR^2}=\left \langle \mathbf{d}\mid\mathbf{d}\right \rangle.
\label{SNR}
\ee
%

\subsection{The impact of the Earth's rotation}
As mentioned in Sec. \ref{methodA},  the term $t_{f}$ in Eq. (\ref{2.8}) represents the effect of the movement of the Earth during the time of the GW signal. If this effect is ignored, $t_{f}$ could be approximately treated as a constant, and the functions $F^+_I$, $F^{\times}_I$ and $\Phi_{IJ}$ are all constants for a given GW event. However, for 3G GW detectors, the cutoff of their low-frequency sensitivity is extended to about 1Hz. Therefore, a detector would observe a given GW event at different time at different location, and it could be treated as a network including a set of detectors at different location with the rotation of the Earth. The baseline is determined by the duration of the GW events. For binary coalescence, the duration of the signal $t_{*}$ is given by \citep{maggiore2008}
\be
	t_{*}=0.86\ \mathrm{day} \left (\frac{1.21\ {M}_{\odot}}{\mathcal{M}_{c}}\right )^{5/3}
	           \left ( \frac{2\ \mathrm{Hz}}{f_{low}}\right )^{8/3},
\ee
where $f_{low}$ is the low-frequency cutoff of the detectors. If we have $f_{low}=1\ \mathrm{Hz}$, for BNSs with $m_{1}=m_{2}=1.4\ {M}_{\odot}$, we have $t_{*}=5.44\ \mathrm{days}$. For SBBHs with $10-10\ {M}_{\odot}$, the result is 4.9 hours and for SBBHs with $30-30\ {M}_{\odot}$, the result is 0.79 hours. Therefore, for the 3G detectors, the impact of the Earth's rotation on the localization of GW sources are important, which includes two effects: One is the modulation of the Doppler effect quantified by the time-dependent function $\Phi_{IJ}$, and the other is quantified by the time-dependent detector responses $F^+_I$ and $F^{\times}_I$. As in the previous work \cite{zhao2018}, in our analysis of this paper, we will also consider these effects when
estimating the localization errors.


 \subsection{The impact of the Earth's revolution}
In this subsection, we will concentrate on the impact of the Earth's revolution, which will be taken into accounted for the first time. There are two effects caused by the revolution of the Earth. The first one is that the angle change affects the azimuth of the detector. So the detector's antenna beam-pattern will change with the Earth's revolution. The solution to this problem is simple. If we replace the Coordinated Universal Time (UTC) with sidereal time, the relative angle between the detectors and the GW source will remain the same at the same time on each sidereal day and the detector's antenna beam-pattern function will not change with the revolution. This effect is equivalent to being absorbed into the the effect of the Earth's rotation. The second one is the Doppler effect caused by the Earth's motion. Since the overall movement speed of the solar system is much smaller than the revolution speed of the Earth, we approximate the Sun as stationary. For convenience, we consider the motion of the Earth in the sun-centered ecliptic coordinate system. Therefore, the time delay $\tau_{e}$ could be written as
\be
	\tau_{e}=\mathbf{n}\cdot(\mathbf{r}_{e}(t_{f})-\mathbf{r}_{e}(t_{c})),
\label{2.21}
\ee
where $\mathbf{r}_{e}$ is the location of the Earth in the Sun-centered ecliptic coordinate system. The smaller $f$, the greater difference between $\mathbf{r}_{e}(t_{f})$ and $\mathbf{r}_{e}(t_{c})$, the more obvious this effect will be.




\section{The Ability of Identifying the Host Galaxy Groups}
\label{nin}
In order to estimate the ability of finding GWs' host groups, for each group in catalog, we assume a GW event of the SBBH merger occurs at its center $(\alpha,\delta,\log({d_{L}}))$. Following \cite{mortonson2008} and \cite{zhao2018b}, we expect measurements to be centered on the central position rather than dispersed by $\sim1\sigma$. These measurements could be seen as the average over many possible observations of the data. The response of GW detector network is given by Eq. (\ref{2.5}). Then, we could get the Fisher matrix and the covariance matrix of source's parameters from Eqs. (\ref{2.17}) and (\ref{2.19}). We derive covariance matrix of the location parameters $\mathbf{Cov}[\alpha,\delta,\log(d_{L})]$ from the whole 9-parameter covariance matrix, and draw an ellipsoid in the parameter space of $(\alpha,\delta,\log(d_{L}))$. In order to simplify the description in the parameter space, we do the following transformation. From the matrix $\mathbf{Cov}[\alpha,\delta,\log(d_{L})]$, we can obtain its eigenvalues $\{\lambda_{1},\lambda_{2},\lambda_{3}\}$, and the corresponding eigenvectors $\{\mathbf{v_{1}},\mathbf{v_{2}},\mathbf{v_{3}}\}$, which could be calculated by using Jacobi eigenvalue algorithm \citep{rutishauser1966}. In the frame with $\{\mathbf{v_{1}},\mathbf{v_{2}},\mathbf{v_{3}}\}$ as the coordinate axes, the covariance matrix is transformed to
\bea
	\mathbf{Cov}'[a,b,c]&=&
	\begin{pmatrix} \mathbf{v_{1}}\\\mathbf{v_{2}}\\\mathbf{v_{3}}\end{pmatrix}
	\mathbf{Cov}[\alpha,\delta,\log(d_{L})]
	\begin{pmatrix} \mathbf{v_{1}}^{T}&\mathbf{v_{2}}^{T}&\mathbf{v_{3}}^{T}\end{pmatrix}\nonumber\\
	&=&\begin{pmatrix}\lambda_{1}&0&0\\0&\lambda_{2}&0\\0&0&\lambda_{3}\end{pmatrix},
\label{3.1}
\eea
where $(a,b,c)$ represent the coordinate in the new coordinate system based on $(\mathbf{v_{1}},\mathbf{v_{2}},\mathbf{v_{3}})$. Therefore, the probability functions of $a$, $b$ and $c$ would be independent. Thus, in the new coordinate system, the possibility density function of the source's location could be simplified as
\be
	f(a,b,c)=C\exp\left (-\frac{1}{2}\left [ \frac{(a-\mu_{a})^{2}}{\lambda_{1}}+\frac{(b-\mu_{b})^{2}}{\lambda_{2}}+\frac{(c-\mu_{c})^{2}}{\lambda_{3}}\right ]\right ),
\label{3.2}
\ee
where $(\mu_{a},\mu_{b},\mu_{c})$ represents the location of the assumed SBBHs and $C$ is a normalized constant. The distribution of $(a-\mu_{a})^{2}/{\lambda_{1}}+(b-\mu_{b})^{2}/{\lambda_{2}}+(c-\mu_{c})^{2}/{\lambda_{3}}$ is a chi-square distribution with three degrees of freedom. Then we can paint an ellipsoid
\be
	\frac{(a-\mu_{a})^{2}}{\lambda_{1}}+\frac{(b-\mu_{b})^{2}}{\lambda_{2}}+\frac{(c-\mu_{c})^{2}}{\lambda_{3}}=\chi^{2},
\label{3.3}
\ee
which stands for the source's possible location given by detectors network. $\chi^{2}$ is a constant related to the confidence we need. Throughout this work, we consider the situation of a 99\% confidence ellipsoid, which follows that $\chi^2=11.34$.

After painting this ellipsoid, we traverse our catalog to find the number of groups in it. The difficulty is that, from GW observation, we can obtain the constraint on the source's luminosity distance $d_L$. However, from the catalog, for each galaxy group, we have only the information of redshift, which relates to $d_L$ by the cosmological model. In particular, for the nearby sources, the relation between them strongly depends on the Hubble constant $H_0$. In this section, we assume the cosmological parameters are all fixed by various observations, and investigate the ability of finding the host galaxy groups from GW observations. In our analysis, the values of $d_{L}$ for the galaxy groups are calculated from the redshift information in catalog by assuming the standard $\Lambda$CDM model with parameters given by WMAP7 \citep{komatsu2011}. The identification ability is quantified by the quantity $N_{in}$, which is the number of galaxy groups in the error ellipsoid. The smaller $N_{in}$ represents the better ability of identification. When $N_{in}=1$, it means that one could find the host galaxy group of this GW event directly without EM counterpart. However, if $N_{in}>1$, it means we cannot find the host group directly from the GW observation alone.

\begin{figure*}
\centering
\subfigure[]{
	\label{omega_a}
	\includegraphics[width=8cm]{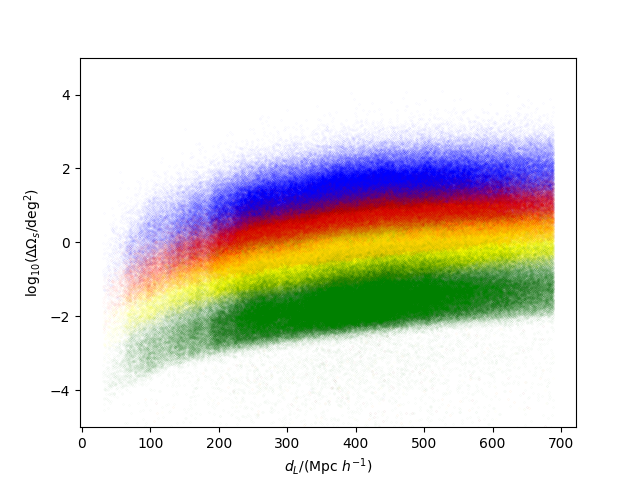}
}
\hspace{5pt}
\subfigure[]{
	\label{omega_b}
	\includegraphics[width=8cm]{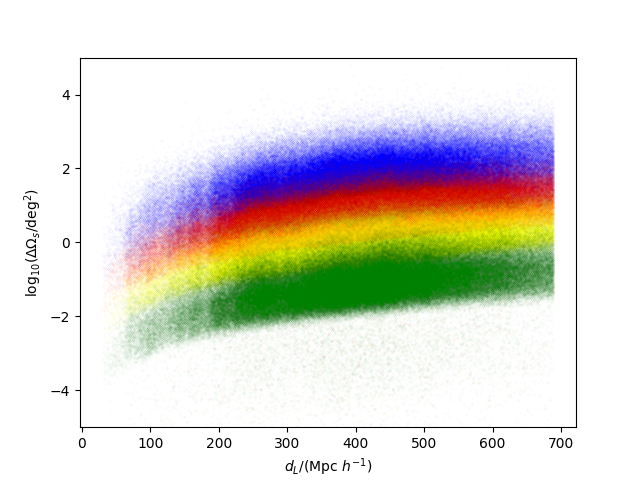}
}

\caption{The distribution of $\Delta\Omega_s$ for different cases. The left panel is for the case with $30-30\ M_{\odot}$ SBBHs and the right panel is for the case with $10-10\ M_{\odot}$ SBBHs. The blue, red, yellow, green dots represent the cases of the LHV, LHVIK, LHVIKCA, CE2ETD networks respectively. Note that, in this figure, the assumed SBBH samples have the random inclination angles. }
\label{figureOmega}
\end{figure*}

\begin{figure*}
\centering
\subfigure[]{
	\label{dL_a}
	\includegraphics[width=8cm]{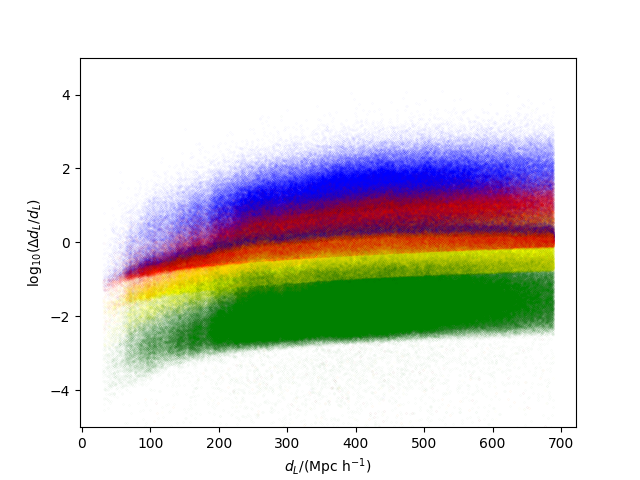}
}
\hspace{5pt}
\subfigure[]{
	\label{dL_b}
	\includegraphics[width=8cm]{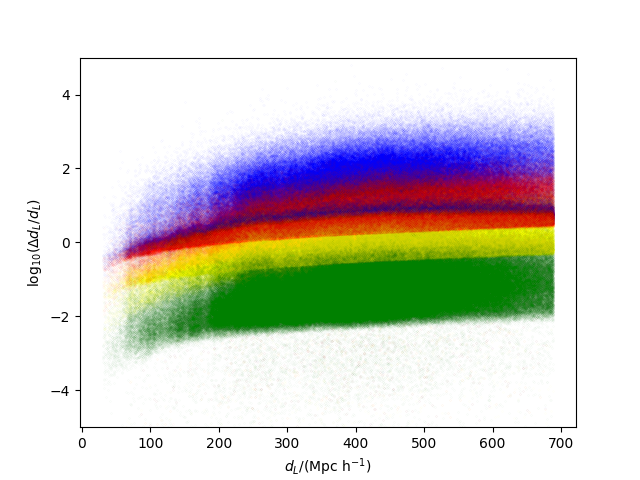}
}

\caption{The distributions of $\Delta d_{L}/d_{L}$ for different cases. The left panel is for the case with $30-30\ M_{\odot}$ SBBHs and the right panel is for the case with $10-10\ M_{\odot}$ SBBHs. The blue, red, yellow, green dots represent the cases of the LHV, LHVIK, LHVIKCA, CE2ETD networks respectively. Note that, in this figure, the assumed SBBH samples have the random inclination angles.}
\label{figuredL}
\end{figure*}

\begin{figure}
\centering
\includegraphics[width=9cm, height=7cm]{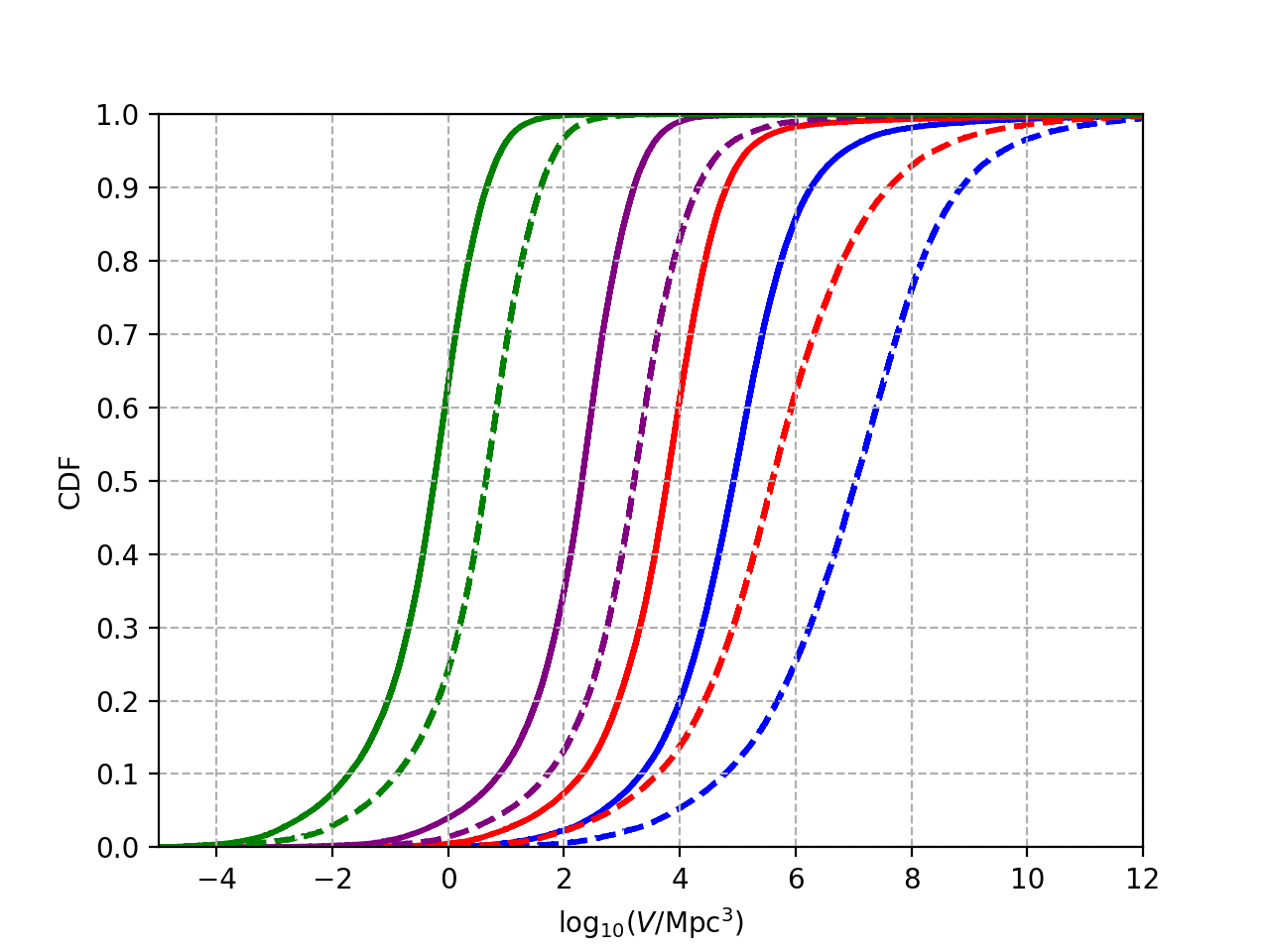}
\caption{The CDF of localization volumes for $30-30\ M_{\odot}$ SBBHs (soild line) and $10-10\ M_{\odot}$ SBBHs (dash line) with different cnetworks. The blue, red, purple and green lines represent the cases of LHV, LHVIK, LHVIKCA, CE2ETD, respectively.}
\label{figureV}
\end{figure}

For a given detector network, the localization ability depends on the feathers of the binary systems, in particular the masses of BHs, and the inclination angle $\iota$ of the systems. For the SBBH mergers in simulation, we assume that their masses are all $30-30\ M_{\odot}$ or $10-10\ M{\odot}$. The distributions of their $\iota$ are all $\propto \sin\iota$. For the detectors network, we consider three different cases for 2G detectors: LHV, LHVIK, LHVIKCA\footnote{L means advanced LIGO-Livingston, H means advanced LIGO-Handford, V means advanced Virgo, I means LIGO-India, K means KAGRA}. Similar to the recent work \cite{sathya2019}, for the 3G detectors, we consider the network consisting of one ET in Europe, one CE in the U.S. and one assumed CE-type detector in Australia. We denote this network as CE2ETD. Note that, for ET, we adopt the sensitivity model referred as ET-D in \cite{punturo2010} and \cite{abernathy2011}. The designed sensitivities of these detectors are given in Fig. \ref{figurenoise}, and the coordinates and orientations of these interferometers are listed in Table \ref{table1}. In this paper, we do not consider the networks of two detectors, since the localization capabilities of the
two-detector networks are very poor \citep{wen2010}. Even for the CE-ETD network, its positioning ability is much worse than the LHVIKCA network \citep{zhao2018}. In Figs. \ref{figureOmega} and \ref{figuredL}, we present the distributions of $\Delta\Omega_{s}$ and $\Delta d_{L}/d_{L}$ with these four networks, which stand for the localization capabilities of various detector networks. In Fig, \ref{figureV}, we show the cumulative distribution function (CDF) of the localization volumes, and find that the typical localization volume for the 2G LHVIK network is around $(10^3-10^5)$Mpc$^{3}$, which can be reduced to around $(0.1-10)$Mpc$^{3}$ in 3G era. Note that, the localization volumes of $30-30\ M_{\odot}$ SBBHs are much smaller compared with \cite{chen2016}, due to the samples' redshift cutoff.


\subsection{The distributions of $N_{in}$}
\begin{figure*}
\centering
\subfigure[LHV]{
	\includegraphics[width=8cm]{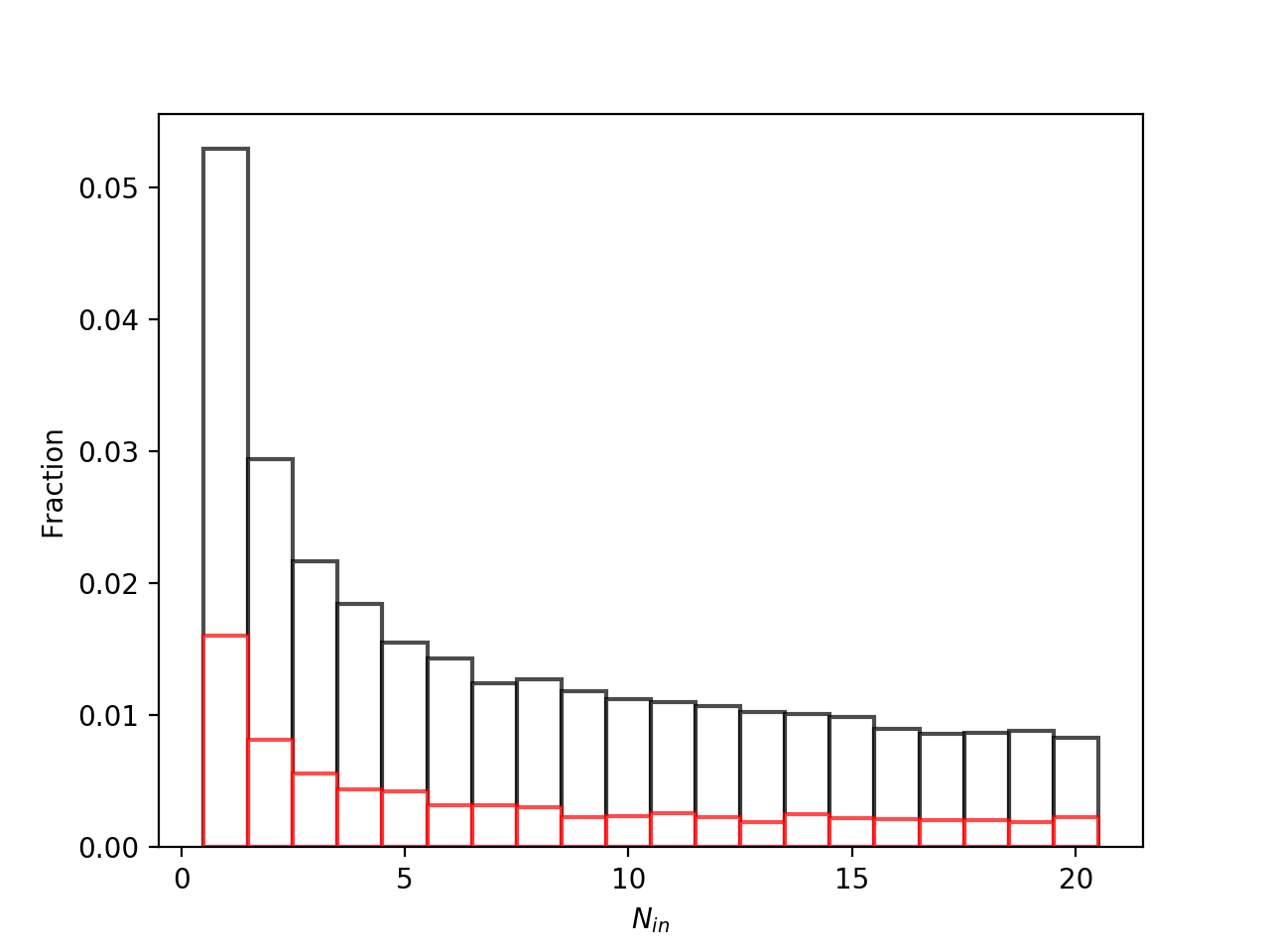}
}
\hspace{2pt}
\subfigure[LHVIK]{
	\includegraphics[width=8cm]{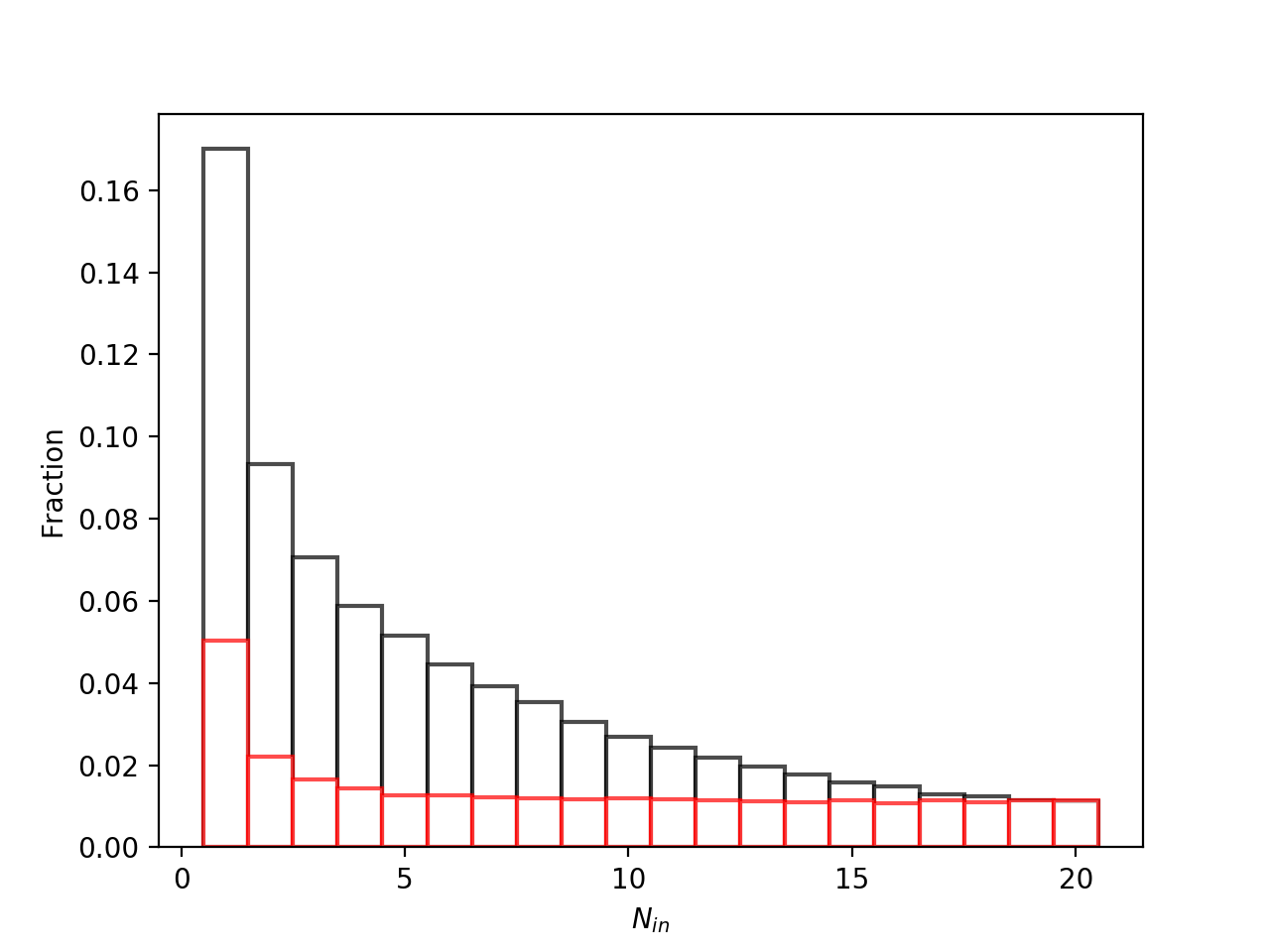}
}
\hspace{2pt}
\subfigure[LHVIKCA]{
	\includegraphics[width=8cm]{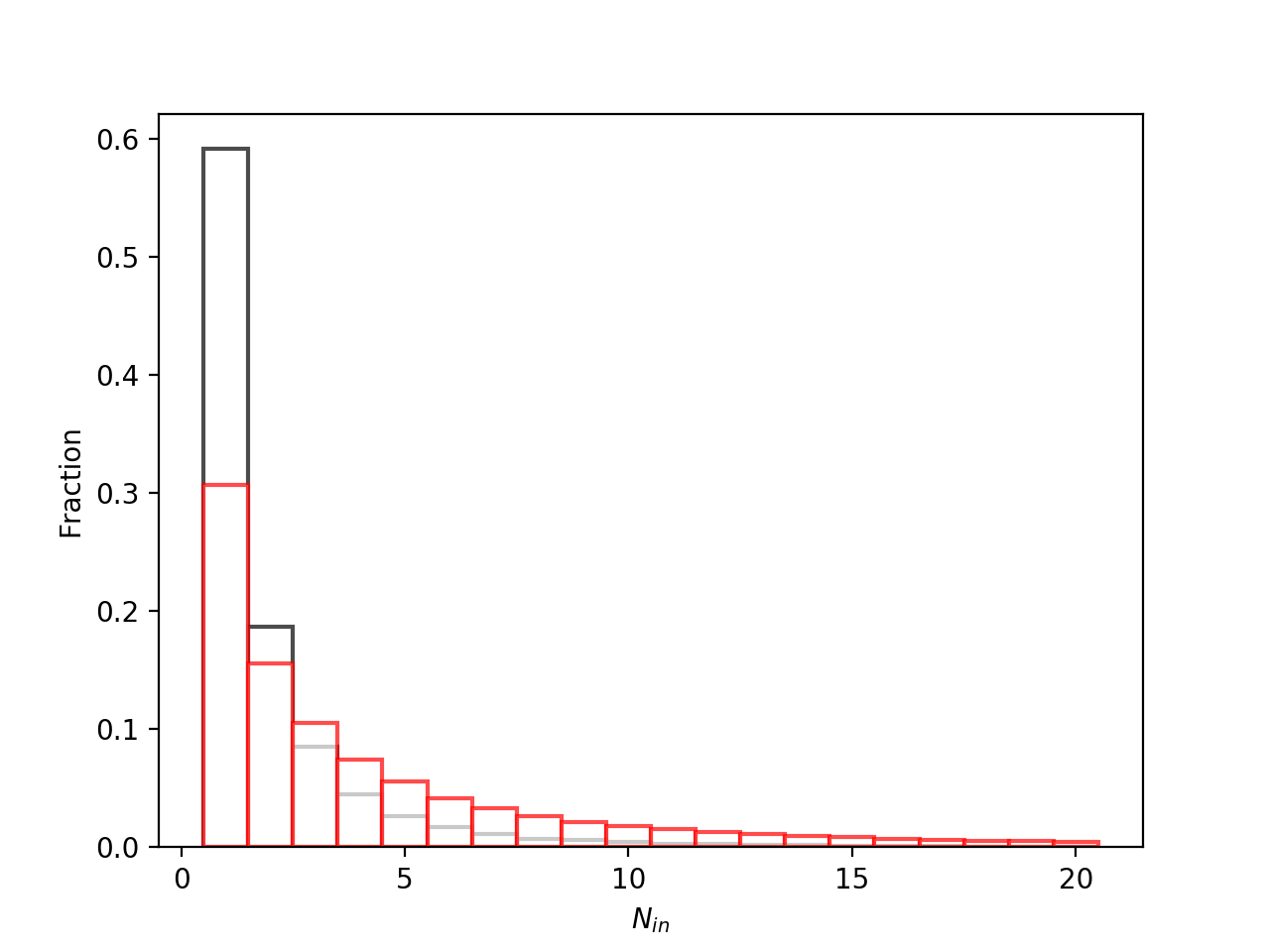}
}
\hspace{2pt}
\subfigure[CE2ETD]{
    \includegraphics[width=8cm]{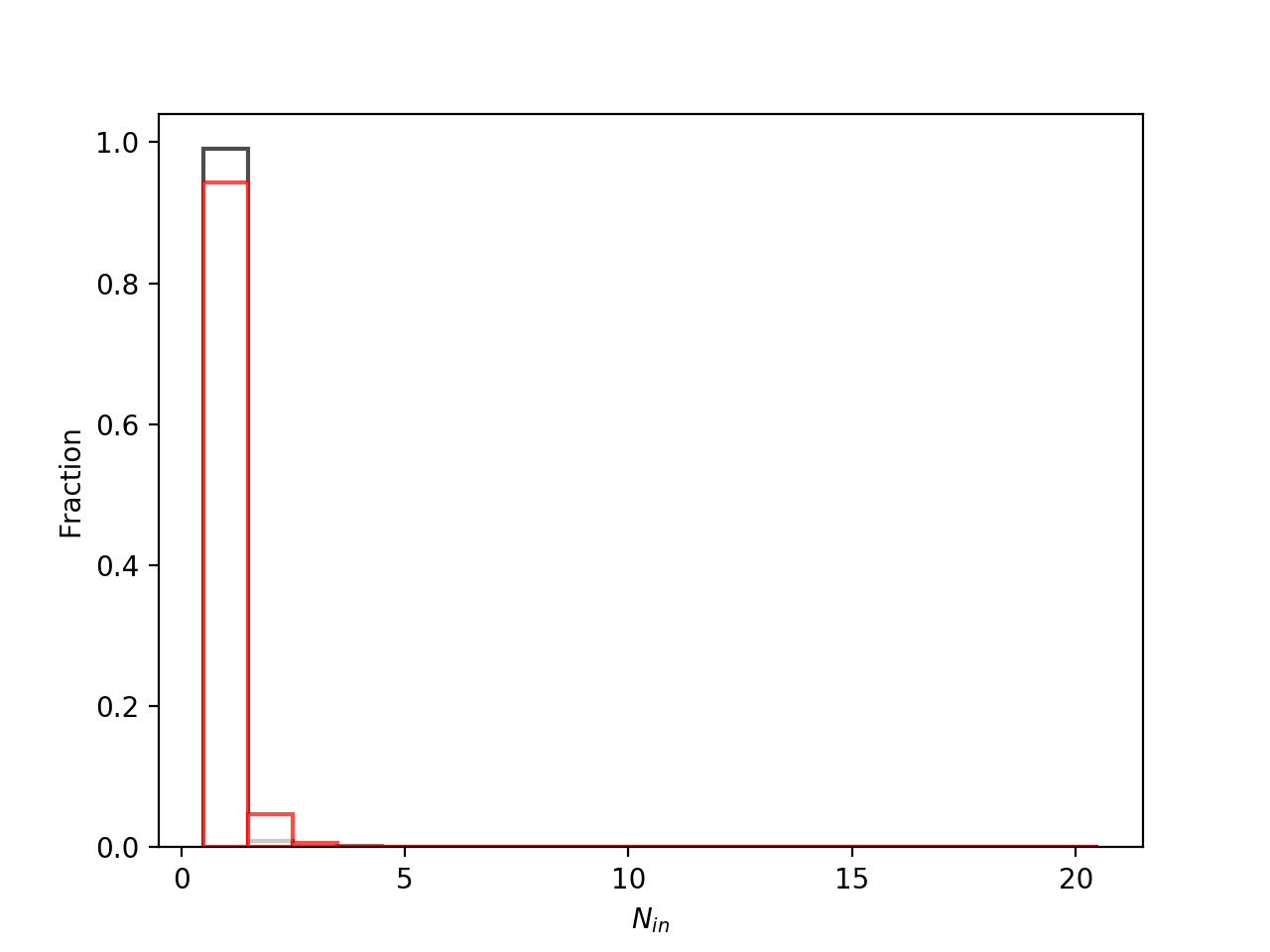}
}
\caption{The probability distribution of $N_{in}$ form 1 to 20 with different networks. In each figure, there are two histograms. The black one is for the distribution with $30-30\ M_{\odot}$ SBBHs and the red one is for that with $10-10\ M_{\odot}$ SBBHs.}
\label{figureLHV}
\end{figure*}

\begin{figure}
\centering
\includegraphics[width=9cm, height=7cm]{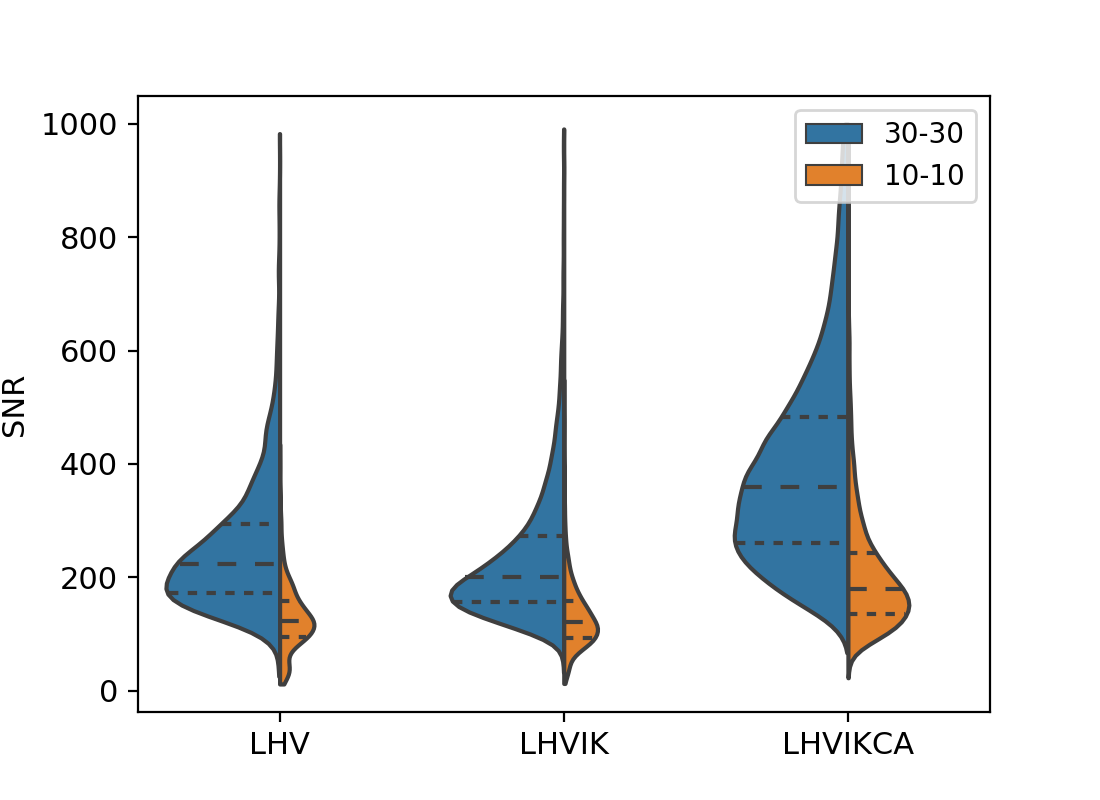}
\caption{The SNR's violin plots of SBBH mergers with $N_{in}=1$. The three violins represent the case of LHV, LHVIK, LHVIKCA network, respectively. In each violin, the left part represents the case of $30-30\ M_{\odot}$ SBBHs and the right part represents the case of $10-10\ M_{\odot}$ SBBHs, while the area of each part is proportional to the SBBHs' number. It is worth noting that the areas between different violins cannot be used for comparison.} 
\label{figureSNR}
\end{figure}

In each case, we calculate $N_{in}$ for the assumed SBBH mergers and get a distribution of $N_{in}$. Fig. \ref{figureLHV} shows the probability distributions of $N_{in}$ from 1 to 20 with different parameters. The integral of these histograms is not necessarily equal to 1, since we have only drawn the part of $N_{in}\le 20$. In each figure, there are two histograms. The black one is for the $30-30\ M_{\odot}$ SBBH samples and the red one is for the $10-10\ M_{\odot}$ SBBH samples. We observe that the heights of the histogram in $30-30\ M_{\odot}$ SBBHs case are higher than that in case with $10-10\ M_{\odot}$ SBBHs. This is not surprising because the cases of $30-30\ M_{\odot}$ SBBHs have stronger GW signals and smaller localization volumes compared with the corresponding cases of $10-10\ M_{\odot}$ SBBHs. There are three violin plots in Fig. \ref{figureSNR} and they represent the SNR distributions of $N_{in}=1$ SBBHs with the LHV, LHVIK, LHVIKCA network. The left part is the result of $30-30\ M_{\odot}$ SBBHs and the right part is the result of $10-10\ M_{\odot}$ SBBHs. The SNRs of $30-30\ M_{\odot}$ SBBHs are much larger than $10-10\ M_{\odot}$ SBBHs'. The fractions of $N_{in}$ with different networks with different parameters are showed in Table \ref{tableNin}, where we list the fractions of $N_{in}=1$, $N_{in}\leq2$, $N_{in}\leq5$, $N_{in}\leq10$ for each case. The rates are also listed under the fractions. The total merger rates can be determined by the current observations of LVC. From the first and second observing runs of LIGO and Virgo, the merger rates of SBBHs are $(9.7-101)\ \mathrm{Gpc}^{-3}\ \mathrm{yr}^{-1}$ at 90\% confidence level \citep{abbott2019a}. So, multiplied by the volume of the redshift (0.01-0.1) area, there would be $(4.0-41)$ SBBH mergers in per year's full-time observations. Since the mass distribution of SBBH mergers are difficult to estimated, we consider two extreme cases, the SBBH mergers' mass are all $30-30\ M_{\odot}$ or all $10-10\ M_{\odot}$ in this paper.

For the LHV network, in the optimal case with $30-30\ M_{\odot}$ SBBHs, the fraction of $N_{in}=1$ is 5.29\%, and the fraction of $N_{in}\leq10$ is 20.05\%. While in the case with $10-10\ M_{\odot}$ SBBHs, the fraction of $N_{in}$ with small value is even less. In the calculation, we find that most of the assumed events have the large positioning areas. So, the LHV network has little probability to identify the host galaxy group of SBBHs.

For the LHVIK network, in the case of $30-30\ M_{\odot}$, the fraction of $N_{in}\leq10$ is 62.05\%, which are much better than the corresponding cases of LHV network. In particular, we find that by GW observations alone, nearly $17\%$ events can identify their host galaxy groups, which would provide the important samples to study the formation circumstances of the SBBHs. However, for the cases with $10-10\ M_{\odot}$ SBBHs, the results 
worse: the fraction of $N_{in}\leq10$ is only 17.62\%, and the fraction of $N_{in}=1$ reduces to $5.04\%$.

From the results of the LHVIKCA network, we observe that if adding two 8-km detectors, the results will be much better. For the case with $30-30\ M_{\odot}$ SBBHs, the fraction of $N_{in}=1$ is 57.19\% and almost all the assumed events have $N_{in}\leq10$. Even for case of $10-10\ M_{\odot}$, nearly $84\%$ assumed events have $N_{in}\leq10$ and 30.75\% of them have $N_{in}=1$.

In comparison with the localization abilities of 2G detector networks, from the results of CE2ETD network, we find the 3G network has a huge improvement. For the $30-30\ M_{\odot}$ SBBHs at low redshift $z<0.1$, almost all events have $N_{in}=1$, i.e. their host galaxy groups can be identified. Even for the $10-10\ M_{\odot}$ SBBHs at the same redshift range, the fraction is also more than $94\%$. These results indicate that, in the 3G era, an abundant SBBH samples with identified host groups can be obtained by GW observations, which are extremely important for the researches on the astrophysical origins of these sources.

\begin{table*}
\begin{center}
\begin{tabular}{c|c|c|c|c|c}
\hline     
     \multicolumn{2}{c|}{$N_{in}$}&1&$\leq2$&$\leq5$&$\leq10$\\
     \hline
     \multirow{2}{*}{\centering LHV}
     &$30-30M_{\odot}$
     &\tabincell{c}{5.29\%\\(0.2-2.2\ yr$^{-1}$)}
     &\tabincell{c}{8.23\%\\(0.3-3.4\ yr$^{-1}$)}
     &\tabincell{c}{13.79\%\\(0.5-5.6\ yr$^{-1}$)}
     &\tabincell{c}{20.05\%\\(0.8-8.2\ yr$^{-1}$)}\\
     \cline{2-6}
     &$10-10M_{\odot}$
     &\tabincell{c}{1.60\%\\(0.06-0.6\ yr$^{-1})$}
     &\tabincell{c}{2.41\%\\(0.09-1.0\ yr$^{-1})$}
     &\tabincell{c}{3.83\%\\(0.2-1.6\ yr$^{-1}$)}
     &\tabincell{c}{5.24\%\\(0.2-2.1\ yr$^{-1}$)}\\
    \hline
     \multirow{2}{*}{LHVIK}
     &$30-30M_{\odot}$
     &\tabincell{c}{17.00\%\\(0.7-6.9\ yr$^{-1}$)}
     &\tabincell{c}{26.33\%\\(1.0-10.7\ yr$^{-1}$)}
     &\tabincell{c}{44.41\%\\(1.7-18.1\ yr$^{-1}$)}
     &\tabincell{c}{62.05\%\\(2.4-25.3\ yr$^{-1}$)}\\
     \cline{2-6}
     &$10-10M_{\odot}$
     &\tabincell{c}{5.04\%\\(0.2-2.0\ yr$^{-1}$)}
     &\tabincell{c}{7.24\%\\(0.3-3.0\ yr$^{-1}$)}
     &\tabincell{c}{11.59\%\\(0.4-4.7\ yr$^{-1}$)}
     &\tabincell{c}{17.62\%\\(0.7-7.2\ yr$^{-1}$)}\\
     \hline
     \multirow{2}{*}{LHVIKCA}
     &$30-30M_{\odot}$
     &\tabincell{c}{59.17\%\\(2.3-24.1\ yr$^{-1}$)}
     &\tabincell{c}{77.89\%\\(3.0-31.8\ yr$^{-1}$)}
     &\tabincell{c}{93.55\%\\(3.7-38.2\ yr$^{-1}$)}
     &\tabincell{c}{98.16\%\\(3.8-40.0\ yr$^{-1}$)}\\
     \cline{2-6}
	 &$10-10M_{\odot}$
	 &\tabincell{c}{30.75\%\\(1.2-12.5\ yr$^{-1}$)}
	 &\tabincell{c}{46.29\%\\(1.8-18.9\ yr$^{-1}$)}
	 &\tabincell{c}{69.86\%\\(2.7-28.5\ yr$^{-1}$)}
	 &\tabincell{c}{83.88\%\\(3.3-34.2\ yr$^{-1}$)}\\
	 \hline
	 \multirow{2}{*}{CE2ETD}
     &$30-30M_{\odot}$
     &\tabincell{c}{99.01\%\\(3.9-40.4\ yr$^{-1}$)}
     &\tabincell{c}{99.92\%\\(3.9-40.8\ yr$^{-1}$)}
     &\tabincell{c}{99.99\%\\(3.9-40.8\ yr$^{-1}$)}
     &\tabincell{c}{100\%\\(3.9-40.8\ yr$^{-1}$)}\\
     \cline{2-6}
     &$10-10M_{\odot}$
     &\tabincell{c}{94.36\%\\(3.7-38.5\ yr$^{-1}$)}
     &\tabincell{c}{99.06\%\\(3.9-40.4\ yr$^{-1}$)}
     &\tabincell{c}{99.93\%\\(3.9-40.8\ yr$^{-1}$)}
     &\tabincell{c}{99.99\%\\(3.9-40.8\ yr$^{-1}$)}\\
     \hline
\end{tabular}
\end{center}
\caption{This table shows the fractions and rates of the assumed GW events with different values of $N_{in}$.}
\label{tableNin}
\end{table*}

\subsection{The edge effect}
As discussed, in our analysis, the constraint of source location is quantified by Eq. (\ref{3.3}). However, our galaxy group catalog covers only a part of the sky area with redshift from 0.01 to 0.1. Therefore, the error ellipsoids of the sources may cover some unobserved regions. So, in this case, the derived values of $N_{in}$ may be smaller than the real value if there are some galaxy groups in the error ellipsoids but outside the considered area. In this subsection, we investigate the impact of this edge effect. In order to quantitatively describe it, in each error ellipsoid derived from the Fisher matrix analysis, we sprinkle 200 points randomly. Then, we get the fraction of the catalog area in the total error ellipsoid by counting the number of the points in the observed region. In this paper, we denote this fraction as $v_{in}$, which quantifies the edge effect mentioned above. If $v_{in}$ is close to 1, the edge effect is negligible for the constraint of this GW event. On the other hand, if $v_{in}$ is much smaller than 1, the edge effect is important, which should be seriously taken into accounted in estimating the localization abilities of GW observations.

In Tables \ref{tableVin}, we list the fraction of the GW events with $v_{in}$ larger than 0.9 in the samples with different values of $N_{in}$ ($N_{in}=1,\ \leq2,\ \leq5,\ \leq10$). We find that almost all the events with small $N_{in}$ have $v_{in}>0.9$, which means that for most SBBH mergers with low $N_{in}$, their error ellipsoids are roughly in the catalog area, and the edge effect is not significant. However, for SBBH mergers with large $N_{in}$, since they are usually accompanied by large location areas, these error ellipsoids usually have a large part outside the catalog areas. Therefore, we conclude that, if we consider only the cases with small $N_{in}$, i.e. $N_{in}\le 10$, the edge effect of the host group localization has little influence on our results.

\begin{table*}
\begin{center}
\begin{tabular}{c|c|c|c|c|c}
\hline     
     \multicolumn{2}{c|}{$N_{in}$}&1&$\leq2$&$\leq5$&$\leq10$\\
     \hline
     \multirow{2}{*}{\centering LHV}
     &$30-30M_{\odot}$
     &\tabincell{c}{92.02\%}
     &\tabincell{c}{91.57\%}
     &\tabincell{c}{90.67\%}
     &\tabincell{c}{89.50\%}\\
     \cline{2-6}
     &$10-10M_{\odot}$
     &\tabincell{c}{91.61\%}
     &\tabincell{c}{91.83\%}
     &\tabincell{c}{90.88\%}
     &\tabincell{c}{90.46\%}\\
    \hline
     \multirow{2}{*}{LHVIK}
     &$30-30M_{\odot}$
     &\tabincell{c}{95.37\%}
     &\tabincell{c}{94.89\%}
     &\tabincell{c}{94.13\%}
     &\tabincell{c}{93.38\%}\\
     \cline{2-6}
     &$10-10M_{\odot}$
     &\tabincell{c}{95.64\%}
     &\tabincell{c}{95.23\%}
     &\tabincell{c}{94.50\%}
     &\tabincell{c}{93.63\%}\\
     \hline
     \multirow{2}{*}{LHVIKCA}
     &$30-30M_{\odot}$
     &\tabincell{c}{97.09\%}
     &\tabincell{c}{96.91\%}
     &\tabincell{c}{96.83\%}
     &\tabincell{c}{96.74\%}\\
     \cline{2-6}
	 &$10-10M_{\odot}$
	 &\tabincell{c}{96.74\%}
	 &\tabincell{c}{96.49\%}
	 &\tabincell{c}{96.83\%}
	 &\tabincell{c}{95.97\%}\\
	 \hline
	 \multirow{2}{*}{CE2ETD}
     &$30-30M_{\odot}$
     &\tabincell{c}{97.34\%}
     &\tabincell{c}{97.35\%}
     &\tabincell{c}{97.35\%}
     &\tabincell{c}{97.35\%}\\
     \cline{2-6}
     &$10-10M_{\odot}$
     &\tabincell{c}{97.06\%}
     &\tabincell{c}{97.04\%}
     &\tabincell{c}{97.02\%}
     &\tabincell{c}{97.02\%}\\
     \hline
\end{tabular}
\end{center}
\caption{This table shows the fractions of SBBHs with $v_{in}>0.9$ in different cases.}
\label{tableVin}
\end{table*}

\section{Measurement of the Hubble Constant}
\label{hubble}

One of the important issues in GW astronomy is that, the GW events produced by the coalescence of compact binaries can act as the standard sirens to constrain the cosmological parameters \citep{schutz1986}. In particular, for a set of the low-redshift GW events, this method is hope to solve the measurement tension of the Hubble constant in the near future (see for instance, \cite{chen2017}). The key role of the standard siren method is to independently determine the redshift of the GW events. For the SBBH events, which lack of EM counterparts, the natural choice is to identify their host galaxies or galaxy groups by the spatial resolutions of GW detector networks, and the redshift information of these GW events can be provided by the optical observations of their hosts. In this section, we will extract the redshift information of GW events from the distribution of galaxy groups in the error ellipsoids by the Bayesian analysis. Then, combining with the luminosity distance uncertainties from GW observation, we will constrain the Hubble constant $H_0$ by using the Hubble's law. Applying the analysis to the 2G and 3G detector networks, we will investigate the potential constraints of $H_0$ by various future GW observations.

\subsection{Bayesian method}
With the GW data $d_{GW}$ and EM data $d_{EM}$ from SDSS DR7 catalog, we use the method given by \citep{chen2017,fishbach2019,soares2019,gray2019} to calculate the posterior probability of $H_{0}$. Different from the
ir measurements, we use the catalog of galaxy groups rather than galaxies in our work. For galaxies, the virial velocity, $v_\mathrm{vir}$, should be considered, which contributes to the main part of the uncertainty of EM data. But for our catalog, the center of each galaxy group is defined to be the stellar mass weighted average of the positions and redshifts of member galaxies so we need not consider the term of $v_\mathrm{vir}$. The uncertainty of term $v_\mathrm{h}$, which is the motion of the halo, still need to consider. \cite{mukherjee2019} gives an estimation of NGC4993 and shows the standard deviation of its halo velocity $\sigma_\mathrm{h}$ is about $80\ \mathrm{km}\ \mathrm{s}^{-1}$. For a group with $z=0.1$, it will cause a 0.27\% uncertainty on its redshift measurement. For most cases, this uncertainty is not important compared with the measurement accuracy of the events' luminosity distances by GW observations, except for a small fraction cases with some low-redshift groups and CE2ETD network. But this fraction is very small compared with the number of full groups in our catalog. For this reason, we will neglect this uncertainty in our calculation. The size of group is another source for the uncertainty of redshift measurement. We find that, in comparison with the uncertainty caused by the motion of galaxy groups, the ratio of the virial radius and distance of a group is generally one order of magnitude smaller. So, we will also neglect this term of uncertainty in our calculation.

      Following the Bayes' theorem, we can write the posterior as
\be
	p(H_{0}|d_{GW},d_{EM})\propto p(d_{GW}, d_{EM}|H_{0})p(H_{0}),
\label{5.1}
\ee
where $d_{GW}$, $d_{EM}$ represent the GW data and EM data respectively, and $p(H_{0})$ represents the prior probability of $H_{0}$. In this work, $d_{GW}$ equals to the GW signals observed by GW detector network and $d_{EM}$ equals to the data from galaxy group catalog. In this section, all the cosmological parameters are fixed, except for $H_{0}$, as the values mentioned in the introduction. In this work, we adopt $H_0=70~\mathrm{km}\ \mathrm{s}^{-1}\ \mathrm{Mpc}^{-1}$ as the fiducial value for the sample simulation. In the Bayesian analysis, we set the prior that the Hubble constant is uniformly distributed in the interval $[60,80]~\mathrm{km}\ \mathrm{s}^{-1}\ \mathrm{Mpc}^{-1}$ in this work.

Following the approach of \cite{chen2017}, the likelihood $p(d_{GW},d_{EM}|H_{0})$ could be written as
\bea
	p(d_{GW},&d_{EM}|H_{0})\propto\frac{1}{\beta(H_{0})}\int p(d_{GW}|d_{L}(z,H_{0}),\Omega)\times\nonumber\\
	&p(d_{EM}|z,\Omega)p_{0}(z,\Omega)d\Omega dz,
\label{5.2}
\eea
where $\beta(H_{0})$ is the normalization term.
Here we could calculate $p(d_{GW}|d_{L}(z,H_{0}),\Omega)$ from Eq. (\ref{3.2}). We ignore the uncertainty of groups' redshift and position. Therefore, $p(d_{EM}|z,\Omega)$ could be written as
\be
	p(d_{EM}|z,\Omega)=\sum_{i}w_{i}\delta(z-z_{i})\delta(\Omega-\Omega_{i}),
\label{5.3}
\ee
where $(z_{i}, \Omega_{i})$ represent the redshift and position of $i$-th possible host group, and $w_{i}$ is the weight of each group, representing the probability of becoming a host group. This weight is related to many factors, such as the stellar mass, morphologies and metallicities (see for instance, \cite{cao2017}). \par

For the group prior $p_{0}(z,\Omega)$, we adopt two different approaches for calculation. In the first approach, we follow the assumption in \cite{soares2019} that the groups are uniformly distributed in comoving volume $V$. So we have
\be
	p_{0}(z, \Omega)\propto\frac{1}{V_{max}}\frac{d^{2}V}{dzd\Omega}\propto\frac{1}{V_{max}}\frac{\chi^{2}(z)}{H(z)},
\label{5.4}
\ee
where $\chi(z)$ is comoving distance and $H(z)$ is Hubble parameter with respect to redshift $z$. In the second approach, we try to get the formalism of $p_{0}(z,\Omega)$ from the group catalog. If assuming the group distribution is isotropic on large scales, we could write $p_{0}(z,\Omega)$ as
\be
	p_{0}(z,\Omega)=p_{0}(z)p_{0}(\Omega)\propto p_{0}(z),
\label{5.5}
\ee
and $p_{0}(z)$ is equivalent to a sum of delta functions at the redshift of the groups in our catalogue since we ignore the uncertainty of groups' redshift. We show the distribution of both group number and stellar mass in Fig. \ref{figuregroup2}. 
In order to ensure the stability of our results, both of these two approaches of $p_0(z,\Omega)$ are discussed in our work. Thus, Eq. (\ref{5.1}) becomes
\be
	p(H_{0}|d_{GW},d_{EM})\propto\frac{p(H_{0})}{\beta(H_{0})}\sum_{i}w_{i}
	p(d_{GW}|d_{L}(z_{i},H_{0}),\Omega_{i})p_{0}(z_{i}, \Omega_{i}).
\label{5.6}
\ee
For $N$ events with sequence numbers $1,2,3,\cdots,j$, this equation could be rewritten as
\bea
	p(H_{0}|d_{GW},d_{EM})\propto\prod_{j=1}^{N}\frac{p(H_{0})}{\beta(H_{0})}
	\times\quad\quad\quad\quad\quad\quad&\nonumber\\\sum_{i}w_{i}
	p(d_{GW,j}|d_{L}((z_{ij},H_{0}),\Omega_{ij})p_{0}(z_{ij},\Omega_{ij}).&
\label{5.7}
\eea
\par

We assume that the data can be detected only if it satisfies some thresholds. For $d_{EM}$, we set its threshold as $z<0.1$. For $d_{GW}$, we set its threshold as $\mathrm{SNR}>12$. So the normalization term $\beta(H_{0})$ is \citep{chen2017,fishbach2019}
\begin{equation}
\label{5.8}
\begin{split}
	\beta(H_{0})&=\int_{d_{EM}>thresh} \int_{d_{GW}>thresh}
	\int p(d_{GW}|d_{L}(z,H_{0}),\Omega)\times\\
	&\quad\quad\quad\quad p(d_{EM}|z,\Omega)p_{0}(z,\Omega)d\Omega dz dd_{EM}dd_{GW},\\
	&=\int P^{GW}_{det}(d_{L}(z,H_{0}),\Omega)P^{EM}_{det}(z,\Omega)p_{0}(z, \Omega)d\Omega dz\\
	&=\int^{z_{h}}_{0}\int P^{GW}_{det}(d_{L}(z,H_{0}),\Omega)p_{0}(z,\Omega)d\Omega dz,
\end{split}
\end{equation}
where we define
\be
	P^{GW}_{det}(d_{L}(z,H_{0}),\Omega) = \int_{d_{GW}>thresh} p(d_{GW}|d_{L}(z,H_{0}),\Omega) dd_{GW},
\label{5.9}
\ee
and
\be
	P^{EM}_{det}(z,\Omega) = \int_{d_{EM}>thresh}p(d_{EM}|z,\Omega) dd_{EM}=\mathcal{H}(z-z_{h}),
\label{5.10}
\ee
where $\mathcal{H}$ is the Heaviside step function. It is worth noting that if the maximum observable distance for the GW events are far less than $z_{h}$, we could neglect the selection effects and set $z_{h}\rightarrow \infty$, and $\beta(H_{0})$ can be simplified to \citep{mandel2016, chen2017, fishbach2019}:
\be
	\beta(H_{0})\propto H_{0}^{3}.
\label{5.11}
\ee
However, in our work, especially for the 3G detector network, the maximum observable distance would be very large and this assumption is no longer appropriate. So, we have to calculate Eq. (\ref{5.8}) numerically. We put an assumed SBBH merger with random $\iota$ in each group. After multiplying the weights $w_{i}$, we can treat the distribution of those assumed SBBH mergers as the term $p(d_{GW}|d_{L}(z,H_{0}),\Omega)$ is Eq. (\ref{5.9}). For each SBBH merger, we calculate its SNR and get the value of $\beta(H_{0})$  from Eq. (\ref{5.8}). We calculate the value of $\beta(H_{0})$ with the prior $H_{0}\in[60,80]~\mathrm{km}\ \mathrm{s}^{-1}\ \mathrm{Mpc}^{-1}$ for different detector networks, and find that for most $H_{0}$ the values of $\beta(H_{0})$ are same since all the mergers' SNR in those cases are larger than 12.



\subsection{Posterior probability of $H_0$ with uniform weight function}

In order to get the constrain of Hubble constant, we select 100 galaxy groups randomly from the catalog, and place a SBBH merger with random $\iota$ at each location. In our calculations, we consider two types of GW sources, i.e. $10-10\ M_{\odot}$ and $30-30\ M_{\odot}$ SBBH mergers with two different choices of prior $p_{0}(z,\Omega)$ as mentioned above. For each event, we can obtain an error ellipsoid in parameter space of $(\alpha,\delta, \log(d_{L}))$ with $99\%$ confidence level from Fisher matrix analysis. Then, we traverse all groups in this ellipsoid, and set their weights $w_{i}=1$ to calculate the posterior probability of $H_0$ by Eq. (\ref{5.7}). Here we need to pay attention to a problem. If the range of the ellipsoid exceeds the maximum redshift $z_{h}$ of the catalog, we may lose some of the possible host galaxy groups outside the catalog, which will induce the bias for the estimation of $H_0$. So, in our analysis, we reject these events to avoid the possible bias. This is treated as the selection criteria for the GW events in our analysis. In the cases with LHV network, we find the values of $\Delta d_{L}/d_{L}$ is very large. Nearly all the GW events cannot satisfy the criteria, and have to be ejected in the analysis. So, we cannot obtain the posterior probability of $H_0$ with this network. For this reason, in this section, we shall only consider the following detector networks: LHVIK, LHVIKCA and CE2ETD.

\begin{figure*}
\centering
\subfigure[LHVIK1]{
    \includegraphics[width=8cm]{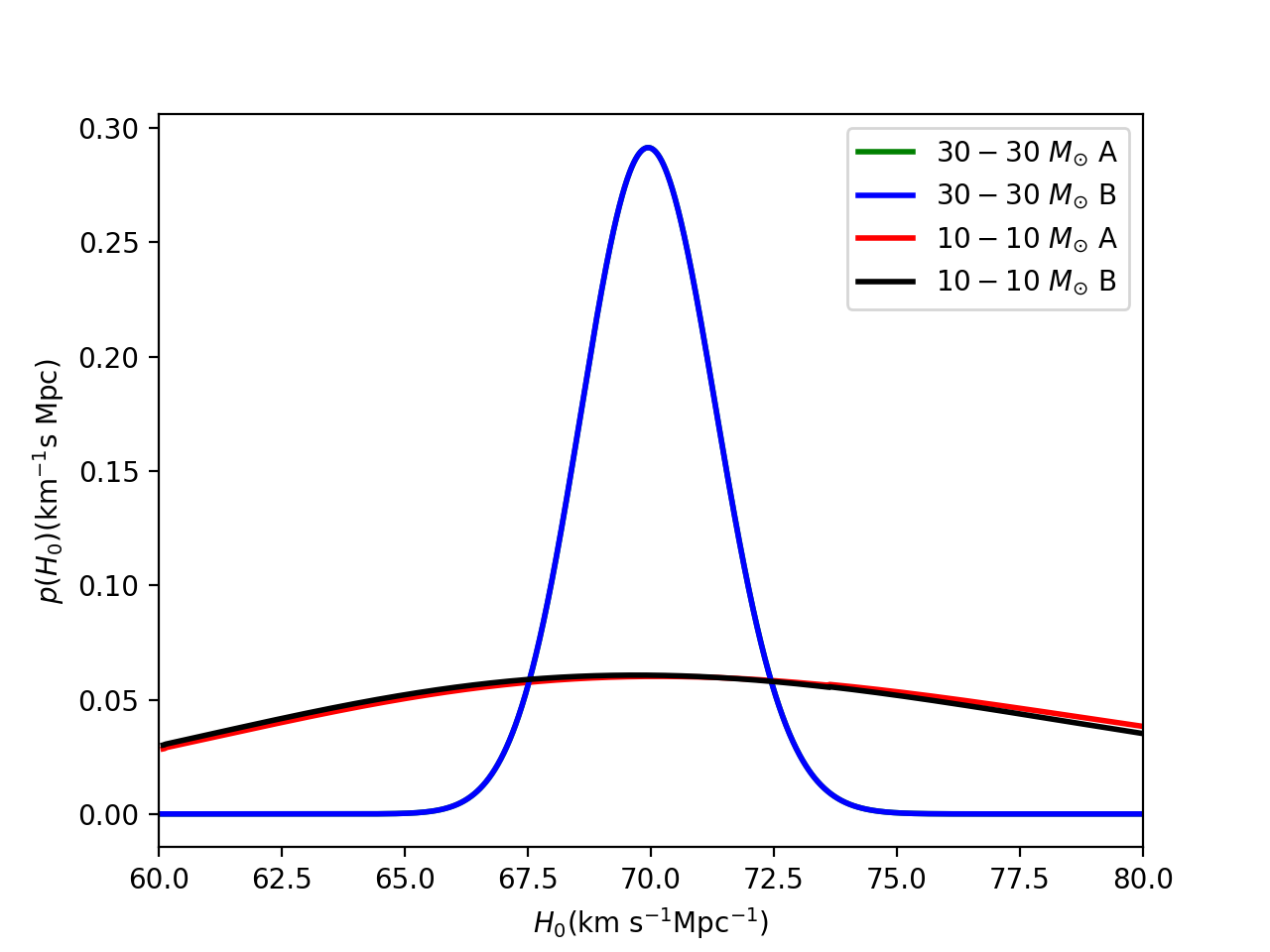}
}
\hspace{2pt}
\subfigure[LHVIK2]{
    \includegraphics[width=8cm]{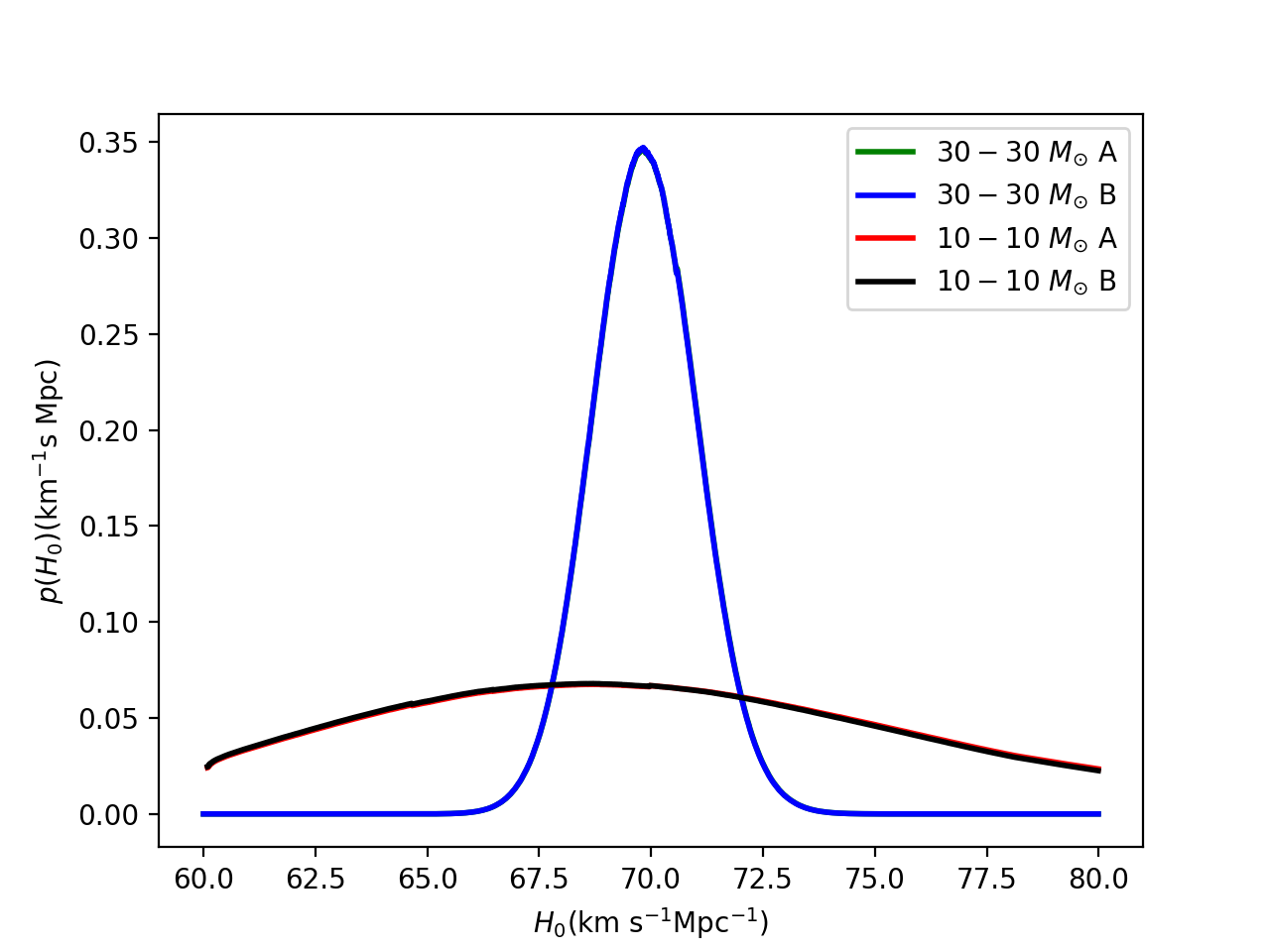}
}
\hspace{2pt}
\subfigure[LHVIKCA1]{
    \includegraphics[width=8cm]{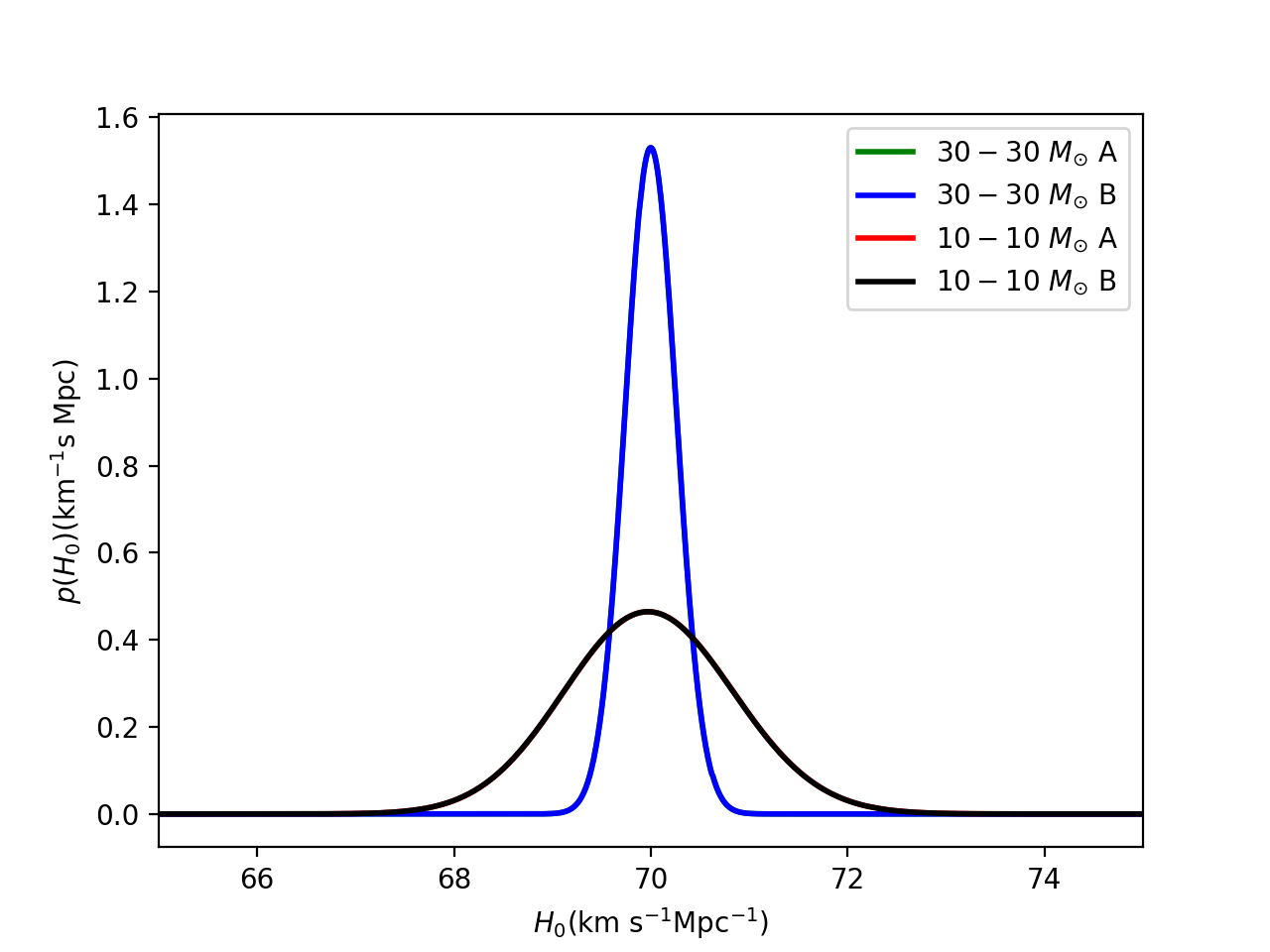}
}
\hspace{2pt}
\subfigure[LHVIKCA2]{
    \includegraphics[width=8cm]{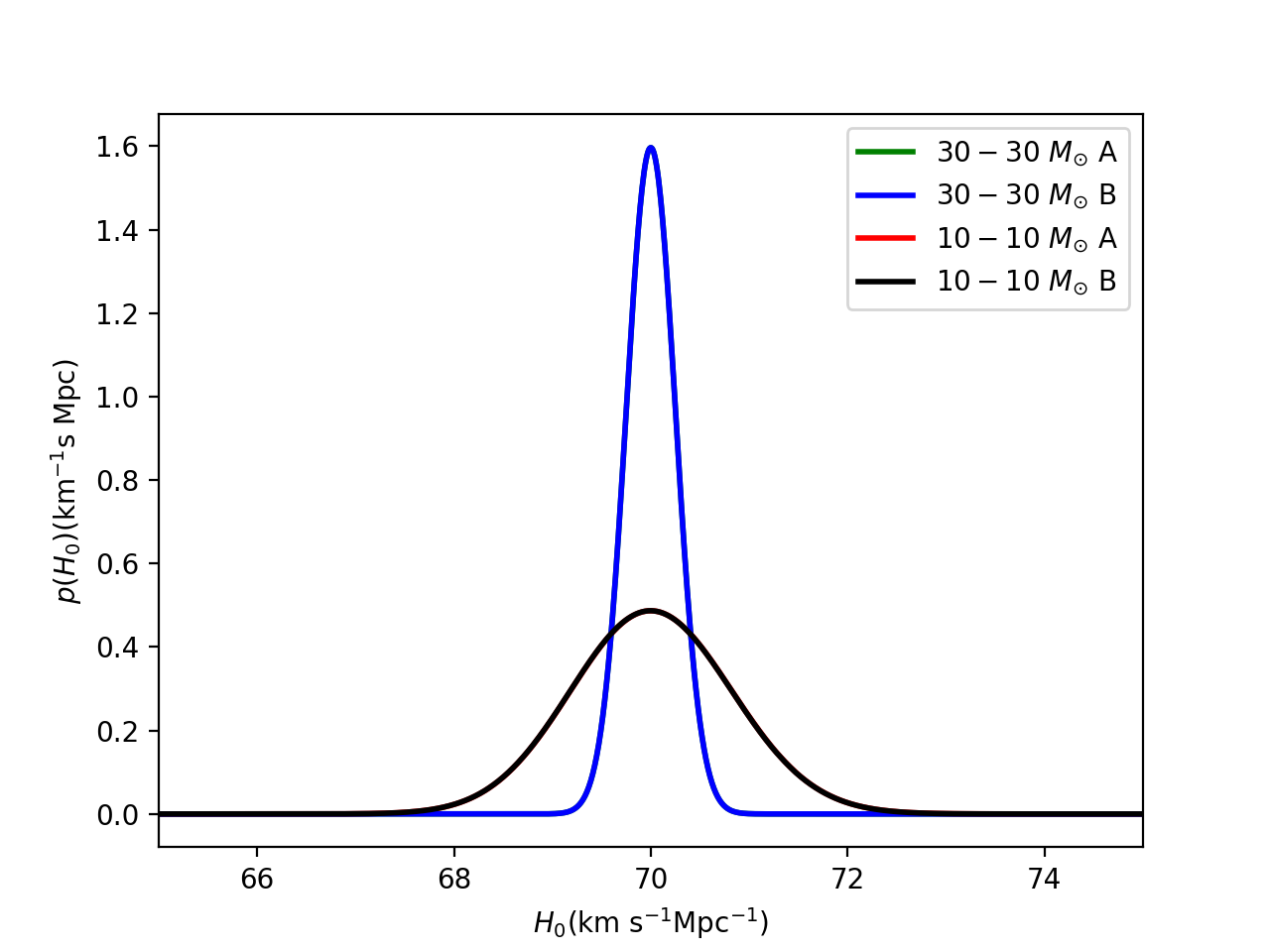}
}
\hspace{2pt}
\subfigure[CE2ETD1]{
    \includegraphics[width=8cm]{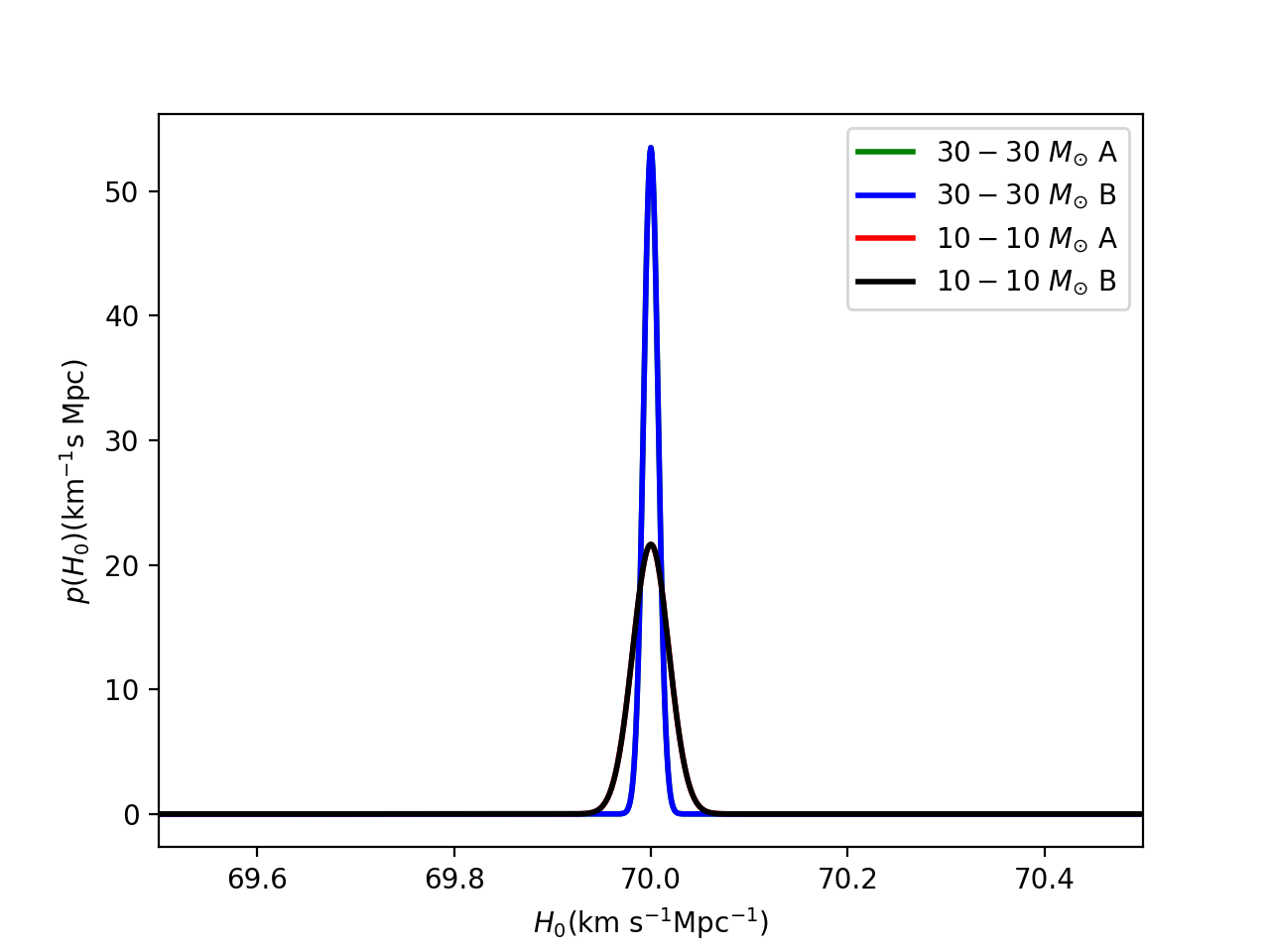}
}
\hspace{2pt}
\subfigure[CE2ETD2]{
    \includegraphics[width=8cm]{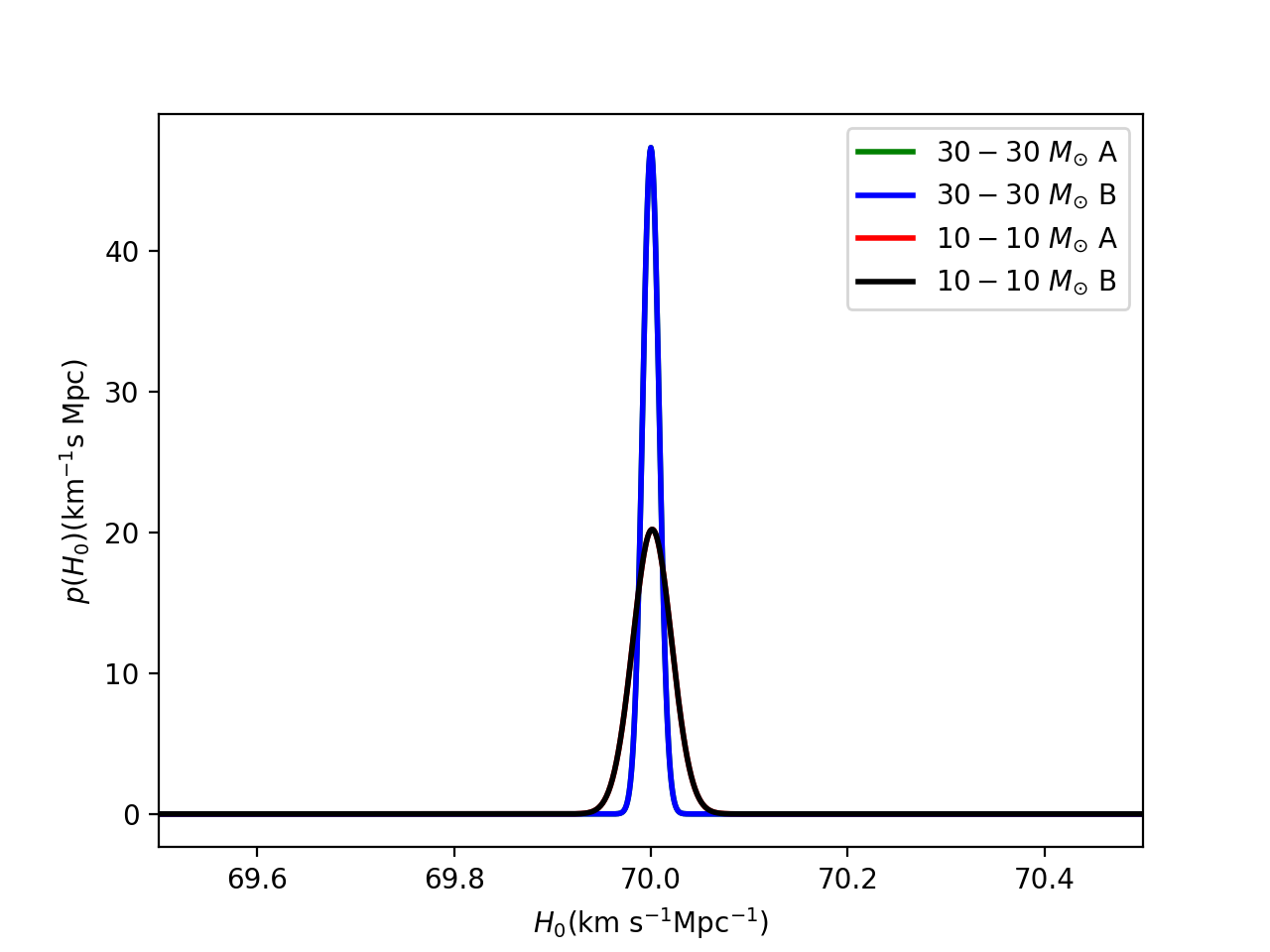}
}
\caption{The posterior distributions of Hubble constant $H_{0}$. The upper, middle and lower panels show the results of for the LHVIK, LHVIKCA, CE2ETD network, respectively. The left panels represent the cases of uniform window function $w_i=1$, and the rights panels represent the cases of $w_{i}$  proportion to the stellar mass of groups. In each panel, there are four lines. The green line is the result in the case of $30-30\ M_{\odot}$ A and the blue one is that of $30-30\ M_{\odot}$ B (see the main text for the details). Similarly, the red line is the result in the case of $10-10\ M_{\odot}$ A, and the black line is that of $10-10\ M_{\odot}$ B. Note that, green line overlaps with blue one, and red line overlaps with black one.}
\label{figureH0}
\end{figure*}

Let us first focus on the 2G LHVIK network. Among the 100 assumed GW events, there are 17 events satisfy the criteria in the case with $30-30\ M_{\odot}$ SBBHs, and only 3 events satisfy the criteria in the case with $10-10\ M_{\odot}$ SBBHs. The posterior distributions of $H_0$ are presented in the upper left panel of Fig. \ref{figureH0}. Note that, in this section, for $30-30\ M_{\odot}$ SBBH mergers, the green curve represents the posterior distribution with uniformly distributed $p_{0}(z,\Omega)$ and we denote it as the case of $30-30\ M_{\odot}$ A. The blue curve represents that with $p_{0}(z,\Omega)$ from group catalog, which is denoted as the case of $30-30\ M_{\odot}$ B. Similarly, for $10-10\ M_{\odot}$ SBBH mergers, the red curve is the posterior with uniformly distributed $p_{0}(z,\Omega)$ and the black curve is the posterior with $p_{0}(z,\Omega)$ from group catalog, denoted as the case of $10-10\ M_{\odot}$ A and the case of $10-10\ M_{\odot}$ B respectively. The posterior to $H_{0}$ is 68.3\% confidence level in this section if without statement. In case of $30-30\ M_{\odot}$ A, the potential constraint is $H_{0}=69.94^{+1.39}_{-1.36}\ \mathrm{km}\ \mathrm{s}^{-1}\ \mathrm{Mpc}^{-1}$, and in the case of $30-30\ M_{\odot}$ B, the constraint is nearly same. So, we find that, different choices of the group prior $p_{0}(z,\Omega)$ cannot significantly affect the constraint of the Hubble constant. On the other hand, for LHVIK network with $10-10\ M_{\odot}$ SBBH mergers, we find the error bars of $H_0$ are quite large in both cases.


For LHVIKCA network with $30-30\ M_{\odot}$ SBBH mergers, we find that 44 events satisfy the criteria and are used for the parameter constraining. The posterior distributions of $H_0$ derived from this network are given in the middle left panel of Fig. \ref{figureH0}, which are $H_0=70.00^{+0.26}_{-0.27}\ \mathrm{km}\ \mathrm{s}^{-1}\ \mathrm{Mpc}^{-1}$ for both case of $30-30\ M_{\odot}$ A and case of $30-30\ M_{\odot}$ B. While for $10-10\ M_{\odot}$ SBBH mergers, we find 21 events can be used for the parameter constraining, and the corresponding result is $H_{0}=69.97^{+0.86}_{-0.85}\ \mathrm{km}\ \mathrm{s}^{-1}\ \mathrm{Mpc}^{-1}$ in the case of $10-10\ M_{\odot}$ A and $H_{0}=69.97\pm0.86\ \mathrm{km}\ \mathrm{s}^{-1}\ \mathrm{Mpc}^{-1}$ in the case of $10-10\ M_{\odot}$ B. Same to the cases of LHVIK network, we find that the effect of group prior $p_{0}(z,\Omega)$ is not significant. In comparison with the corresponding cases with LHVIK network, we find the uncertainties of $H_0$ are nearly five times smaller. Due to the contribution of the two 8-km detectors in the network, the localization ability for the individual event is significantly improved, which leads to two effects on the $H_0$ constraining: First, the uncertainties of the redshift distribution of the GW event is reduced. Second, the number of the GW events satisfying the criteria is increased. So, we conclude that, if the proposals of 8-km detectors can be realized in the near future, the capabilities of GW cosmological, based on the SBBHs as the standard sirens, will be greatly improved in the 2G detector era.

Now, we turn to the 3G detector network CE2ETD. As anticipated, we find that the accuracies of $H_0$ are much better than that of the 2G detector networks. For $30-30\ M_{\odot}$ mergers, 66 events can be used for the parameter constraining, which follow the result $H_{0}=70.000\pm0.008\ \mathrm{km}\ \mathrm{s}^{-1}\ \mathrm{Mpc}^{-1}$ for both cases with different prior choices. For $10-10\ M_{\odot}$ SBBH mergers, the constraint of $H_{0}$ is $70.000\pm0.018\ \mathrm{km}\ \mathrm{s}^{-1}\ \mathrm{Mpc}^{-1}$, derived from 65 events, which satisfy the selection criteria. The posterior distributions in these cases are presented in the lower left panel of Fig. \ref{figureH0}. We find that, even comparing with the results of 2G LHVIKCA detector network, the uncertainty of $H_0$ is reduced by a factor of $30$. \cite{nair2018} discussed a single ET's accuracy of $H_{0}$ measurement. Our results are obviously better, since the localization volumes for the CE2ETD network are much smaller than a single ET's.

For a given detector network, the potential constraint on the Hubble constant depends also on the event rate of the SBBHs. If the statistical uncertainties are dominant, the uncertainty of $H_{0}$ is proportional to $1/\sqrt{N}$, where $N$ is the number of GW events. If assuming all GW events of SBBHs are $30-30\ M_{\odot}$ systems, for LHVIK network, the uncertainty of $H_{0}$ is about $(1.0-3.1)\ \mathrm{km}\ \mathrm{s}^{-1}\ \mathrm{Mpc}^{-1}$ in five years' observation. For LHVIKCA network, the value reduces to $(0.18-0.59)\ \mathrm{km}\ \mathrm{s}^{-1}\ \mathrm{Mpc}^{-1}$, and for CE2ETD network it becomes $(0.006-0.018)\ \mathrm{km}\ \mathrm{s}^{-1}\ \mathrm{Mpc}^{-1}$. However, if assuming all the SBBH GW events are $10-10\ M_{\odot}$ systems, the constraints of $H_0$ become slightly looser. For the LHVIKCA network, we have $\Delta H_0=(0.60-1.92)\ \mathrm{km}\ \mathrm{s}^{-1}\ \mathrm{Mpc}^{-1}$, and for the CE2ETD network, it is $\Delta H_0=(0.013-0.040)\ \mathrm{km}\ \mathrm{s}^{-1}\ \mathrm{Mpc}^{-1}$. These results are all listed in the upper part of Table \ref{tableH0}, which indicate that, by identifying the host galaxy groups, the SBBH GW events have the ability to follow the tight constraints on Hubble constant.

Note that, in the previous work \cite{chen2017}, the authors carried out the similar analysis to constrain the Hubble constant by SBBH standard sirens. The difference is that, in this paper, the authors tried to determine the redshifts of GW events by identifying their host galaxies, rather than the galaxy groups. The authors predicted a 20\% measurement of $H_{0}$ with $30-30\ M_{\odot}$ SBBH mergers and a 10\% measurement with $10-10\ M_{\odot}$ SBBH mergers by 2026, where the potential observations contain one year's LHV O3 run, two years' LHV at design sensitivity and two years' LHVIK at design sensitivity. In comparison with these results, we find our results with LHVIK network are better, which is mainly caused by the following facts: (1) In the optimal case, we assume all SBBHs have the same mass, which could overestimate the merger rate $30-30\ M_{\odot}$ SBBH mergers. (2) As mentioned above, in our calculations, we assume the network run all the observation time.

\begin{table*}
\centering
\begin{tabular}{c|c|c|c|c }
	\hline
	Network & BH mass&f$_\mathrm{effective}$&N effective events&Five-year observation\\
	\hline
	LHVIK & $30-30\ M_{\odot}$&17\%&~8.1\%/$\sqrt{N}$&1.4\%-4.4\%\\
	LHVIK & $10-10\ M_{\odot}$&3\%&NA&NA\\
	LHVIKCA & $30-30\ M_{\odot}$&44\%&~2.5\%/$\sqrt{N}$&0.26\%-0.85\%\\
	LHVIKCA & $10-10\ M_{\odot}$&21\%&~5.6\%/$\sqrt{N}$&0.86\%-2.74\%\\
	CE2ETD & $30-30\ M_{\odot}$&66\%&~0.09\%/$\sqrt{N}$&~0.008\%-0.026\%\\
	CE2ETD & $10-10\ M_{\odot}$&65\%&~0.21\%/$\sqrt{N}$&~0.018\%-0.057\%
	\\
	\hline
	LHVIK & $30-30\ M_{\odot}$&22\%&~7.7\%/$\sqrt{N}$&1.2\%-3.7\%\\
	LHVIK & $10-10\ M_{\odot}$&5\%&NA&NA\\
	LHVIKCA & $30-30\ M_{\odot}$&37\%&~2.1\%/$\sqrt{N}$&0.24\%-0.78\%\\
	LHVIKCA & $10-10\ M_{\odot}$&24\%&~5.7\%/$\sqrt{N}$&0.82\%-2.62\%\\
	CE2ETD & $30-30\ M_{\odot}$&63\%&~0.09\%/$\sqrt{N}$&~0.008\%-0.026\%\\
	CE2ETD & $10-10\ M_{\odot}$&61\%&~0.22\%/$\sqrt{N}$&~0.020\%-0.064\%
	\\
	\hline
\end{tabular}
\caption{
The results of constraining the Hubble constant with different networks. The third row is the fraction of effective events in the 100 random samples. The fourth row lists the results of $\Delta H_0/H_0$ as functions of the number of effective events $N$. The last row means the results of $\Delta H_0/H_0$ for five years' GW observations. Note that, in upper part and lower part we consider the uniform window function $w_i=1$ and $w_{i}$ proportional to the stellar mass of groups, respectively.}
\label{tableH0}
\end{table*}


\subsection{Effect of weight functions}

Generally, the event rate of SBBH mergers is related to the rate of star formation, which depends on the stellar mass, star forming gas metallicity, etc. In \cite{cao2017}, a comprehensive analysis of host galaxy properties of SBBH mergers was presented by implementing semi-analytical recipes for SBBH formation and merger into cosmological galaxy formation model, and we can observe that there is a significant relationship between the rate of events and the galaxies' stellar mass. In this subsection, we will investigate the influence of the weight function $w_i$, which describes the possibility of the a SBBH merger occurs in this galaxy group during the observation time, on the constraining of Hubble constant. As a rough estimation, in the sample simulation and Bayesian analysis, we use the stellar mass of group to represent the weight $w_i$ and get the posterior probability of Hubble constant with 100 samples again.
We show our results in the right panels of Fig. \ref{figureH0} and in the lower part of Table \ref{tableH0}. Similar to the part of $w_{i}=1$, there are four lines in each figure to represent four different cases as mentioned above. From Table \ref{tableH0}, we find that for all the cases with different detector networks, the results are all very close to the corresponding cases with uniform weight function. For instance, if SBBH mergers are all $30-30\ M_{\odot}$ systems, for LHVIK network, the uncertainty of Hubble constant is $\Delta H_0=(0.8-2.6)\ \mathrm{km}\ \mathrm{s}^{-1}\ \mathrm{Mpc}^{-1}$ in five years' observation. For LHVIKCA network, the result is $\Delta H_0=(0.17-0.55)\ \mathrm{km}\ \mathrm{s}^{-1}\ \mathrm{Mpc}^{-1}$, and for CE2ETD network, it is $\Delta H_0=(0.006-0.018)\ \mathrm{km}\ \mathrm{s}^{-1}\ \mathrm{Mpc}^{-1}$. If SBBH mergers are all $10-10\ M_{\odot}$ systems, we have $\Delta H_0=(0.57-1.83)\ \mathrm{km}\ \mathrm{s}^{-1}\ \mathrm{Mpc}^{-1}$ for LHVIKCA network, and $\Delta H_0=(0.014-0.045)\ \mathrm{km}\ \mathrm{s}^{-1}\ \mathrm{Mpc}^{-1}$ for CE2ETD network.

In order to stabilize the constraints of $H_0$ derived above, we have also considered another case for cross-checking: We seed the SBBHs randomly in groups as done by setting $w_i=1$ in Sec. 5.2, but estimate $\Delta H_0$ by assuming $w_i$ is proportional to group stellar mass. We find that, for each detector network, the potential constraint of $H_0$ is nearly same with those in the two cases discussed above.

\section{Conclusions}
\label{conclusion}
The detection of GW signals of the compact binary coalescence by LVC opens a new window to understand our universe. From the GW waveforms of these events, we could obtain the sources' luminosity distances, which are independent of the cosmic distance ladder. If their redshift information can also be obtained from observations, this kind of GW events can act as the standard sirens to constrain various cosmological parameters, including Hubble constant, dark energy and so on. The reshifts of GW events can be generally derived from the optical observation on their host galaxies, groups, or clusters. In addition, investigation on the properties of host galaxies can also help us to understand the formation mechanism of these compact binaries. Therefore, identifying the hosts of GW events is the key role for these issues. For the coalescing BNSs, such as GW170817, we could identify their EM counterparts directly by observing their EM signals in various frequency channels. However, for SBBH mergers, we could hardly find their EM counterparts. So, we should search for other ways to identify their hosts. One of the well studied methods is to localize the host candidates of compact binaries by utilizing the spatial resolutions of the GW detector networks, and then identify them by comparing with the galaxy catalogs. In most previous works, the authors tried to identify their host galaxies, and found they are hard to be uniquely determined.

In this paper, we discuss the feasibility to identify the SBBH mergers' host groups, rather than the host galaxies, with help of the spatial resolution of GW detector networks. In the current scenario of structure formation in standard cosmological model, galaxies were formed and reside in dark matter halos. Therefore, in comparison with the galaxy, the halo-based galaxy group stands for the larger physical structure in the universe. For a given GW event, its host galaxy group is much easier, than the host galaxy, to be identified due to the larger spatial size. In addition, identification of host galaxy groups can partly overcome the problems due to the incompleteness of galaxy catalog, and that due to the peculiar velocities of host galaxies in the dark matter halos. 
In our analysis, as an example of application of the method, we consider the galaxy group catalog with redshift interval $z\in(0.01,0.1)$ derived from SDSS DR7 data. In each galaxy group in the catalog, we place an assumed SBBH merger. Then, we use the Fisher matrix technique to investigate the localization abilities of GW detector networks. For each assumed merger, we count the number of groups $N_{in}$ in its localization area to quantify the localization capabilities of various detector networks, including the LHV, LHVIK, LHVIKCA networks in 2G era, and the CE2ETD network in 3G era. For each detector network, we consider two kinds of GW sources with different masses, i.e. $30-30\ M_{\odot}$ SBBHs and $10-10\ M_{\odot}$ SBBHs. We find that, for the LHV network, it seems difficult to identify the host galaxy group of SBBHs. However, for the LHVIK network, the situation becomes much better: For case with $30-30\ M_{\odot}$ SBBHs, the host galaxy groups of about $17\%$ events in this redshift range can be identified, and for the case with $10-10\ M_{\odot}$ SBBHs, this fraction is about $5\%$. In particular, if we consider the contribution of two proposed 8-km detectors in the 2G network, the fraction is increased to $\sim 60\%$ for the $30-30\ M_{\odot}$ SBBHs, and $\sim 30\%$ for the $10-10\ M_{\odot}$ SBBHs. We also investigate the identification ability of 3G detector network CE2ETD, which consists of one ET detector in Europe, and two CE-type detectors in the U.S and Australia respectively, and find that the host galaxy groups of about $99\%$ $30-30\ M_{\odot}$ SBBHs can be identified their host groups, while for the $10-10\ M_{\odot}$ SBBHs, this fraction is about $94\%$.

For each SBBH merger observed by the detector network, from the redshift distribution of the group candidates, which are in the error ellipsoid of the GW event, we can extract the redshift information of this event. Combining with the measurement of the luminosity distance $d_L$ from the GW observation, we investigate the potential constraints on the Hubble constant $H_0$ by various 2G and 3G detectors. For each detector network, we consider two extreme cases for the masses of the binaries, i.e. in the first case, we assume all the SBBHs are $30-30\ M_{\odot}$ binaries, and in the second case, we assume that all the SBBHs are $10-10\ M_{\odot}$ binaries. Considering the event rate of SBBH mergers derived from the current observations of LVC, we calculate the potential constraint $\Delta H_0$ by the Bayesian analysis. We find that five-year's full time observation of 2G detector network LHVIK can follow a constraint of $\Delta H_0/H_0\sim (1.4\%, 4.4\%)$ in the optimal case of $30-30\ M_{\odot}$ SBBHs. However, in the other extreme case of $10-10\ M_{\odot}$ SBBHs, the potential constraint of $H_0$ is too loose. If we take into account the two 8-km detectors, we have the constraint of $\Delta H_0/H_0\sim (0.26\%, 0.85\%)$ for the extreme case with only $30-30\ M_{\odot}$ SBBHs, and $\Delta H_0/H_0\sim (0.86\%, 2.74\%)$ for the case with only $10-10\ M_{\odot}$ SBBHs. In the 3G era, if considering five-year observations of CE2ETD detector network , we expect to get $\Delta H_0/H_0\sim (0.008\%, 0.026\%)$ for the case with only $30-30\ M_{\odot}$ SBBHs, and $\Delta H_0/H_0\sim (0.018\%, 0.057\%)$ for the case with only $10-10\ M_{\odot}$ SBBHs.

In \cite{vallisneri2007}, \cite{grover2013} and \cite{Rodriguez2013}, the validity of the Fisher matrix is discussed. Fisher information matrix seems to have accuracy problems in the prediction of observation parameters for 2G detectors, especially for low SNR detection. In our work, we choose $z=0.1$ as the redshift cutoff, and only discuss the cases of a network of at least three 2G detectors, which ensures that most of our samples have a high SNR, especially for those with low $N_{in}$. We find about 96\% $10-10\ M_{\odot}$ SBBH samples with the LHV network and $N_{in}\leq20$ have SNR larger than 50. In the follow-up work, we will consider using more accurate simulation methods to give predictions of observation parameters.

At the end of this paper, we should emphasize that although we have considered the SDSS DR7 group catalog as an example of application, the similar analysis can be directly applied to other galaxy surveys, including DESI (Dark Energy Spectroscopic Instrument), Euclid, LSST (Large Synoptic Survey Telescope), CSST (China Space Station Telescope) and so on.

\section*{Acknowledgements}

We appreciate the helpful discussions with Xiaohu Yang, Bin Hu, David Blair, Linqing Wen and Xian Chen. This work is supported by NSFC Grants No. 11773028, No. 11633001, No. 11653002, No. 11421303, No. 11903030, No. 11903033, No. 11690024, the Fundamental Research Funds for the Central Universities, the China Postdoctoral Science Foundation Grant No. 2019M652193, the Strategic Priority Research Program of the Chinese Academy of Sciences Grant No. XDB23010200 and No. XDB23040100 and the 973 Project Grant No. 2015CB857000.

\section*{DATA AVAILABILITY}
The data underlying this article will be shared on reasonable request to the corresponding author.




\begin{thebibliography}{99}
\bibitem[\protect\citeauthoryear{Abazajian et al.}{2009}]{abazajian2009}Abazajian, K.~N., Adelman-McCarthy, J.~K., Ag${\rm\ddot{u}}$eros, M.~A., et al.\ 2009, \apjs, 182, 543
\bibitem[\protect\citeauthoryear{Abbott et al.}{2016a}]{abbott2016a}Abbott, B.~P., Abbott, R., Abbott, T.~D., et al.\ 2016, \prl, 116, 061102
\bibitem[\protect\citeauthoryear{Abbott et al.}{2016b}]{abbott2016b}Abbott, B.~P., Abbott, R., Abbott, T.~D., et al.\ 2016, \prl, 116, 241103
\bibitem[\protect\citeauthoryear{Abbott et al.}{2016c}]{abbott2016c}Abbott, B.~P., Abbott, R., Abbott, T.~D., et al.\ 2016, \prd, 93, 122003
\bibitem[\protect\citeauthoryear{Abbott et al.}{2016d}]{abbott2016d}Abbott, B.~P., Abbott, R., Abbott, T.~D., et al.\ 2016,
\prx, 6, 041015
\bibitem[\protect\citeauthoryear{Abbott et al.}{2016f}]{LIGOA+}Abbott, B.~P., et al., {\emph{The lsc-virgo white paper on instrument science}} (2016-2017 edition), Technical Report No. T1400226, LIGO and Virgo Scientific Collaborations, LIGO Internal Document No. T1400226, 2016
\bibitem[\protect\citeauthoryear{Abbott et al.}{2017a}]{abbott2017a}Abbott, B.~P., Abbott, R., Abbott, T.~D., et al.\ 2017,
\prl, 118, 221101
\bibitem[\protect\citeauthoryear{Abbott et al.}{2017b}]{abbott2017b}Abbott, B.~P., Abbott, R., Abbott, T.~D., et al.\ 2017,
\prl, 119, 141101
\bibitem[\protect\citeauthoryear{Abbott et al.}{2017c}]{abbott2017c}Abbott, B.~P., Abbott, R., Abbott, T.~D., et al.\ 2017,
\prl, 119, 161101
\bibitem[\protect\citeauthoryear{Abbott et al.}{2017d}]{abbott2017d}Abbott, B.~P., Abbott, R., Abbott, T.~D., et al.\ 2017,
\apjl, 848, L13
\bibitem[\protect\citeauthoryear{Abbott et al.}{2017e}]{abbott2017e}Abbott, B.~P., Abbott, R., Abbott, T.~D., et al.\ 2017,
\apjl, 851, L35
\bibitem[\protect\citeauthoryear{Abbott et al.}{2017f}]{abbott2017f}Abbott, B.~P., Abbott, R., Abbott, T.~D., et al.\ 2017,
\cqg, 34, 044001
\bibitem[\protect\citeauthoryear{Abbott et al.}{2018a}]{abbott2018a}Abbott, B.~P., Abbott, R., Abbott, T.~D., et al.\ 2018,
Living Reviews in Relativity, 21, 3
\bibitem[\protect\citeauthoryear{Abbott et al.}{2019a}]{abbott2019a}Abbott, B.~P., Abbott, R.,  Abbott, T.~D., et al.\ 2019,
\prx, 9, 031040
\bibitem[\protect\citeauthoryear{Abbott et al.}{2019b}]{abbott2019b}Abbott, B.~P., Abbott, R.,  Abbott, T.~D., et al.\ 2019, arXiv preprint arXiv:1908.06060
\bibitem[\protect\citeauthoryear{Abernathy et al.}{2011}]{abernathy2011}Abernathy, M., Acernese, F., Ajith, P., et al. Einstein gravitational wave Telescope conceptual design study[J]. 2011.
\bibitem[\protect\citeauthoryear{Adhikari et al.}{2020}]{adhikari2020}Adhikari, S., Fishbach, M., Holz, D.~E., et al.\ 2020, arXiv preprint arXiv:2001.01025
\bibitem[\protect\citeauthoryear{Ajith et al.}{2008}]{ajith2008}Ajith, P., Babak, S., Chen, Y., et al.\ 2008, \prd, 77, 104017
\bibitem[\protect\citeauthoryear{Arcavi et al.}{2017}]{arcavi2017}Arcavi, I., Hosseinzadeh, G., Howell, D.~A., et al.\ 2017,
 \nat, 551, 64
\bibitem[\protect\citeauthoryear{Artale et al.}{2019}]{artale2019}Artale, M.~C., Mapelli, M., Giacobbo, N., et al.\ 2019, \mnras, 487, 1675
\bibitem[\protect\citeauthoryear{Artale et al.}{2020}]{artale2020}Artale, M.~C., Bouffanais, Y., Mapelli, M., et al.\ 2020, \mnras, 495, 1841

\bibitem[\protect\citeauthoryear{Bell et al.}{2003}]{Bell et al.(2003)}Bell, E. F., McIntosh, D. H., Katz, N., Weinberg, M. D., 2003, \apjs,
149, 289
\bibitem[\protect\citeauthoryear{Blair et al.}{2015}]{blair2015}Blair, D., Ju, L., Zhao, C.~N., et al.\ 2015, Science China Physics, Mechanics \& Astronomy, 58, 120402
%
\bibitem[\protect\citeauthoryear{Blanton et al.}{2003a}]{blanton2003a}Blanton, M. R., et al.\ 2003a, \apj, 592, 819
\bibitem[\protect\citeauthoryear{Blanton et al.}{2003b}]{blanton2003b}Blanton, M. R., et al.\ 2003a, Astron. J., 125, 2348
\bibitem[\protect\citeauthoryear{Blanton et al.}{2007}]{blanton2007}Blanton, M. R., \& Roweis, S.\ 2007, Astron. J., 133, 734
\bibitem[\protect\citeauthoryear{Borch et al.}{2006}]{borch2006}Borch, A., Meisenheimer, K., Bell, E. F., Rix, H.-W., Wolf, C., Dye, S., Kleinheinrich, M., \& Kovacs, Z.\ 2006, \aanda, 453, 869


\bibitem[\protect\citeauthoryear{Cao et al.}{2017}]{cao2017} Cao, L., Lu, Y., Zhao, Y.\ 2017, \mnrad, 474, 4997
\bibitem[\protect\citeauthoryear{Chen \& Holz}{2016}]{chen2016}Chen, H.~Y., \& Holz, D.~E.\ 2016, arXiv preprint arXiv:1612.01471
\bibitem[\protect\citeauthoryear{Chen et al.}{2018}]{chen2017}Chen, H.~Y., Fishbach, M., Holz, D.~E., \ 2018, Nature, 526, 545
\bibitem[\protect\citeauthoryear{Coulter et al.}{2017}]{coulter2017}Coulter, D.~A., Foley, R.~J., Kilpatrick, C.~D., et al.\ 2017, Science, 358, 1556
\bibitem[\protect\citeauthoryear{Cramer}{1989}]{cramer1989}Cramer, H.\ 1989 \emph{Mathematical Methods of Statistics (Princeton University, Princeton, 1946)}, Google Scholar
\bibitem[\protect\citeauthoryear{Cutler \& Flanagan}{1994}]{cutler1994}Cutler, C., Flanagan, E.~E.\ 1994, \prd, 49, 2658
\bibitem[\protect\citeauthoryear{Del Pozzo}{2012}]{pozzo2012}Del Pozzo, W.\ 2012, \prd, 86, 043011
\bibitem[\protect\citeauthoryear{Dotti et al.}{2006}]{dotti2006}Dotti, M., Salvaterra, R., Sesana, A., et al.\ 2006, \mnrad, 372, 869-875
\bibitem[\protect\citeauthoryear{Dwyer et al.}{2015}]{dwyer2015}Dwyer, S., Sigg, D., Ballmer, S.~W., et al.\ 2015, \prd, 91, 082001
\bibitem[\protect\citeauthoryear{Farr}{2019}]{farr2019}Farr, W.~M., Fishbach, M., Ye, J., et al.\ 2019, \apjl, 883, L42
\bibitem[\protect\citeauthoryear{Finn}{1992}]{finn1992}Finn, L.~S.\ 1992, \prd, 46, 5236
\bibitem[\protect\citeauthoryear{Finn \& Chernoff}{1993}]{finn1993}Finn, L.~S., Chernoff, D.~F.\ 1993, \prd, 47, 2198
\bibitem[\protect\citeauthoryear{Fishbach et al.}{2019}]{fishbach2019}Fishbach, M., Gray, R., Hernandez, I.~M., et al.\ 2019, \apjl, 871, L13
\bibitem[\protect\citeauthoryear{Goldstein et al.}{2017}]{goldstein2017}Goldstein, A., Veres, P., Burns, E., et al.\ 2017, \apjl, 848, L14
\bibitem[Gray et al.(2019)]{gray2019}Gray, R., Hernandez, I.~M., Qi, H., et al.\ 2019, arXiv preprint arXiv:1908.06050
\bibitem[\protect\citeauthoryear{Grover et al.}{2013}]{grover2013}Grover, K., Farihurst, S., Farr, B. F., et al.\ 2013, \prd, 89, 042004
\bibitem[\protect\citeauthoryear{Gupta et al.}{2019}]{sathya2019}Gupta, A., Fox, D., Sathyaprakash, B. S., Schutz, B. F., 2019, \apj, 886, 71

\bibitem[\protect\citeauthoryear{Howell et al.}{2017}]{howell2017}Howell, E.~J., Chan, M.~L., Chu, Q., et al.\ 2017, \mnrad, 474, 4385-4395
\bibitem[\protect\citeauthoryear{Jaranowski et al.}{1998}]{jaranowski1998}Jaranowski, P., Krolak, A., Schutz, B.~F.\ 1998, \prd, 58, 063001
\bibitem[\protect\citeauthoryear{Kroupa}{2001}]{kroupa2001}Kroupa, P.\ 2001, \mnrad, 322, 231
\bibitem[\protect\citeauthoryear{Komatsu et al.}{2011}]{komatsu2011}Komatsu, E., Smith, K.~M., Dunkley, J., et al.\ 2011, \apjs, 192, 18
\bibitem[\protect\citeauthoryear{LIGO Collaboration et al.}{2017}]{ligo2017}LIGO Scientific Collaboration, Virgo Collaboration, 1M2H Collaboration, et al.\ 2017, \nat, 551, 85-88
\bibitem[\protect\citeauthoryear{Macleod \& Hogan}{2008}]{macleod2008}Macleod, C. L., \& Hogan, C. J.\ 2008, \prd, 77, 043512
\bibitem[\protect\citeauthoryear{Maggiore}{2008}]{maggiore2008} Maggiore, M.\ 2008, \emph{Gravitational Waves, Volume 1: Theory and Experiments}, Oxford University Press
\bibitem[\protect\citeauthoryear{Mandel et al.}{2016}]{mandel2016}Mandel, I., Farr, W., Gair, J. 2016, Tech. Rep. P1600187, LIGO
%
\bibitem[\protect\citeauthoryear{Mapelli}{2016}]{Mapelli2016}Mapelli, M., 2016, \mnrad, 459, 3432
\bibitem[\protect\citeauthoryear{Messenger \& Read}{2012}]{messenger2012}Messenger, C., Read, J.\ 2012, \prl, 108, 091101
\bibitem[\protect\citeauthoryear{McKernan et al.}{2018}]{McKernan2018}Mckernan, B., Ford, K. E. S., Bellovary, J., et al. 2018, \apj, 866, 66
\bibitem[\protect\citeauthoryear{Moesta et al.}{2012}]{moesta2012}Moesta, P., Alic, D., Rezzolla, L., et al.\ 2012, \apjl, 749, L32
\bibitem[\protect\citeauthoryear{Mortonson \& Hu}{2008}]{mortonson2008}Mortonson, M.~J., \& Hu, W.\ 2008, \prd, 77, 043506
\bibitem[\protect\citeauthoryear{Mosta et al.}{2010}]{mosta2010}Mosta, P., Palenzuela, C., Rezzolla, L., et al.\ 2010, \prd, 81, 064017
\bibitem[\protect\citeauthoryear{Mukherjee \& Wandelt}{2018}]{mukherjee2018}Mukherjee, S., Wandelt, B.~D.\ 2018, arXiv preprint arXiv:1808.06615
\bibitem[\protect\citeauthoryear{Mukherjee et al.}{2019}]{mukherjee2019}Mukherjee, S., Lavaux, G., Bouchet, F.~R., et al,\ 2019, arXiv preprint arXiv:1909.08627
\bibitem[\protect\citeauthoryear{Nair et al.}{2018}]{nair2018}Nair, R., Bose, S., \& Siani, T.~D.\ 2018, \prd, 98, 023502
\bibitem[\protect\citeauthoryear{Nishizawa}{2017}]{ni2017}Nishizawa A.\ 2017, \prd, 96, 101303.
\bibitem[\protect\citeauthoryear{Oguri}{2016}]{oguri2016}Oguri, M.\ 2016, \prd, 93, 083511
\bibitem[\protect\citeauthoryear{Palmese et al.}{2019}]{palmese2019}Palmese, A., Graur, O., Annis, J.~T., et al.\ 2019,  arXiv preprint arXiv:1903.04730
\bibitem[\protect\citeauthoryear{Petrosian}{1976}]{petrosian1976}Petrosian, V.\ 1976, \apj, 209, L1
\bibitem[\protect\citeauthoryear{Punturo et al.}{2010}]{punturo2010}Punturo, M., Abernathy, M., Acernese, F., et al.\ 2010, \cqg, 27, 194002
%
\bibitem[\protect\citeauthoryear{Rodriguez et al.}{2013}]{Rodriguez2013}Rodriguez, C. L., Farr, B., Farr, W. M. \& Mandel I.\ 2013, \prd, 88, 084013
\bibitem[\protect\citeauthoryear{Rodriguez et al.}{2015}]{Rodriguez2015}Rodriguez, C. L., Morscher, M., Pattabiraman, B., et al. 2015, \prl, 115, 051101
%
%
\bibitem[\protect\citeauthoryear{Rodriguez et al.}{2016}]{Rodriguez2016}Rodriguez, C. L., Chatterjee, S., Rasio, F. A., 2016, \prd, 93, 084029
%
%
%
%
%
\bibitem[\protect\citeauthoryear{Rutishauser}{1966}]{rutishauser1966}Rutishauser, H.\ 1966 Numerische Mathematik, 9, 1-10
\bibitem[\protect\citeauthoryear{Savchenko et al.}{2017}]{savchenko2017}Savchenko, V., Ferrigno, C., Kuulkers, E., et al.\ 2017, \apjl, 848, L15
\bibitem[\protect\citeauthoryear{Sasaki et al.}{2016}]{Sasaki2016}Sasaki, M., Suyama, T., Tanaka, T., Yokoyama, S., 2016, \prl, 117, 061101
\bibitem[\protect\citeauthoryear{Sathyaprakash et al.}{2010}]{Sathya2010}Sathyaprakash, B. S., Schutz, B., van den Broeck, C., \cqg, 27, 215006
%
\bibitem[\protect\citeauthoryear{Schlegel et al.}{1998}]{schlegel1998}Schlegel, D. J., Finkbeiner, D. P., Davis, M.\ 1998, \apj, 500, 525
\bibitem[\protect\citeauthoryear{Schneider et al.}{2017}]{Schneider2017}Schneider, R., Graziani, L., Marassi, S., et al. 2017, \mnrad, 471, L105
\bibitem[\protect\citeauthoryear{Schnittman}{2011}]{schnittman2011}Schnittman, J.~D.,\ 2011, \cqg, 28, 094021
\bibitem[\protect\citeauthoryear{Schutz}{1986}]{schutz1986}Schutz, B.~F.,\ 1986, \nat, 323, 310.
\bibitem[\protect\citeauthoryear{Soares-Santos et al.}{2017}]{soares2017}Soares-Santos, M., Holz, D.~E., Annis, J., et al.\ 2017, \apjl, 848, L16
\bibitem[\protect\citeauthoryear{Soares-Santos et al.}{2019}]{soares2019}Soares-Santos, M., Palmese, A., Hartley, W., et al.\ 2019, arXiv preprint arXiv:1901.01540
\bibitem[\protect\citeauthoryear{Strauss et al.}{2002}]{strauss2002}Strauss, M. A., et al.\ 2002, Astron. J., 124, 1810
\bibitem[\protect\citeauthoryear{Tabvir et al.}{2017}]{tanvir2017}Tanvir, N.~R., Levan, Astron. J., Gonzalez-Fernandez, C., et al.\ 2017, \apjl, 848, L27
\bibitem[\protect\citeauthoryear{Taylor \& Gair}{2012}]{taylor2012}Taylor, S.~R., Gair, J.~R.\ 2012, \prd, 86, 023502
\bibitem[\protect\citeauthoryear{Unnikrishnan et al.}{2013}]{unni2013}Unnikrishnan, C.~S.,\ 2013, \ijmpd, 22, 1341010
\bibitem[\protect\citeauthoryear{Valenti et al.}{2017}]{valenti2017}Valenti, S., David, J., Yang, S., et al.\ 2017, \apjl, 848, L24
\bibitem[\protect\citeauthoryear{Vallisneri}{2007}]{vallisneri2007}Vallisneri, M.\ 2007, \prd, 77, 042001

\bibitem[\protect\citeauthoryear{Vitale \& Evans}{2017}]{vitale2017}Vitale, S., Evans, M.\ 2017, \prd, 95, 064052
%
\bibitem[\protect\citeauthoryear{Wang et al.}{2020}]{wang2020}Wang, B., Zhu, Z. Y., Li, A., Zhao, W., 2020, arXiv preprint arXiv:2005.12875
%
%
\bibitem[\protect\citeauthoryear{Wang et al.}{2016}]{Wang2016}Wang, L., Spurzem, R., Aarseth, S., et al., 2016, \mnrad, 458, 1450
%
%
%
\bibitem[\protect\citeauthoryear{Wang et al.}{2018}]{Wang2018}Wang, S., Wang, Y. F., Huang, Q. G., \& Li, T. G. F., \prl, 120, 191102
\bibitem[\protect\citeauthoryear{Wen \& Chen}{2010}]{wen2010}Wen, L., Chen, Y.\ 2010, \prd, 81, 082001
\bibitem[\protect\citeauthoryear{Yan et al.}{2020}]{yan2020}Yan, C. S., Zhao, W., \& Lu, Y. J., \ 2020, \apj, 889, 79
\bibitem[\protect\citeauthoryear{Yang et al.}{2005}]{yang2005}Yang, X., Mo, H. J., Jing, Y. P., van den Bosch, F. C.\ 2005a, \mnrad, 358, 217
\bibitem[\protect\citeauthoryear{Yang et al.}{2007}]{yang2007}Yang, X., Mo, H.~J., van den Bosch, F.~C., et al.\ 2007, \apj, 671, 153
\bibitem[\protect\citeauthoryear{Zhao et al.}{2011}]{zhao2011}Zhao, W., van den Broeck, C., Baskaran, D., Li, T. G. F.,\ 2011, \prd, 83, 023005
\bibitem[\protect\citeauthoryear{Zhao \& Santos}{2019}]{zhao2017}Zhao, W., Santos, L.\ 2019, \jcap, 11, 009
\bibitem[\protect\citeauthoryear{Zhao \& Wen}{2018}]{zhao2018}Zhao, W., Wen, L.\ 2018, \prd, 97, 064031
\bibitem[\protect\citeauthoryear{Zhao et al.}{2018}]{zhao2018b}Zhao, W., Wright, B.~., \& Li, B.\ 2018, \jcap, 10, 052
\bibitem[\protect\citeauthoryear{Zhang et al.}{2017}]{zhang2017}Zhang, X., Yu, J., Liu, T., Zhao, W., Wang, A. Z., \ 2017, \prd, 95, 124008
\end{thebibliography}




\bsp	
\label{lastpage}
\end{document}